# Optical Nanoresonators

V.V.Klimov
Lebedev Physical Institute
E-mail: klimov256@gmail.com

Translated by A.V. Sharonova

The review presents an analysis and generalization of classical and most modern approaches to the description and development of operation of open optical nanoresonators, that is, resonators all sizes of which are smaller than the resonant wavelength of radiation in a vacuum. Particular attention is paid to the physics of such phenomena as bound states in a continuum, anapole states, supercavity modes, and perfect nonradiating modes with extremely high quality factors and localizations of electromagnetic fields. An analysis of the optical properties of natural oscillations in nanoresonators made of metamaterials is also presented in the review. The effects considered in this review, besides being of fundamental import can also find applications in the development of optical nanoantennas, nanolasers, biosensors, photovoltaic devices, and nonlinear nanophotonics.

1.Introduction

2. General Approaches to Describing the Properties of Optical Nanoresonators

2.1. Closed Resonator Eigenmodes

2.2. Eigenmodes of Open Nanoresonators

    *2.2.1. What are open nanoresonator modes?*

    *2.2.2. Frequency as mode eigenvalue*

    *2.2.3. Permittivity as mode eigenvalue*

    *2.2.4. Perfect nonradiating modes*

# 1. Introduction

For modern optical devices, the localization of light in volumes of all sizes smaller than the wavelength in vacuum, that is in nanoresonators, becomes critically important. The localization is extremely important for such modern applications of nano-optics as lasers and spasers, optical nano-antennas, metasurfaces, chemical and biological sensors, optical computers, filters, switches, detectors, optical memory elements. The compactness of these devices allows one to integrate them into optical microcircuits, and that is extremely important since the miniaturization process in the production of electronic microcircuits has almost reached the theoretical limit. In this regard, the resonant optical properties of plasmonic nanoparticles [1–6] and dielectric nanoparticles with a high refractive index [7–25] are being actively studied currently. Figure 1 shows examples of optical nanoresonators.

The physics of optical phenomena in such nanostructures is very complex and leads to many interesting applications, such as nanoantennas [5,6,8-14], plasmonic [2,4] and non-plasmon nanolasers [22, 25,26], and non-linear nanophotonics [17, 21, 29, 30]. As in any other field of physics, here all effects are associated with the existence of certain modes of natural oscillations in nanoparticles.

For applications, eigenmodes with strong field localization and low radiative losses are of particular interest. Modes of this kind have recently attracted the closest attention of leading scientific groups, which have discovered several types of weakly radiating systems: bound states in a continuum [27–30], anapole current distributions [19, 31–36], supercavity modes [25, 37, 38], perfect nonradiating modes [39-41].

Due to radiative losses, intermode interference and amplification of both electric and magnetic fields, the physics of high-$Q$ modes in nanocavities is very complex. In particular, the fields of usual quasi-normal modes increase without limit at infinity. In view of the exceptional importance of open nanoresonators, a number of approaches are being intensively developed at present for their description and creation of new nanodevices based on them. These approaches are not well known to a wide range of physicists working in the field of nano-optics and nanoplasmonics and their applications.

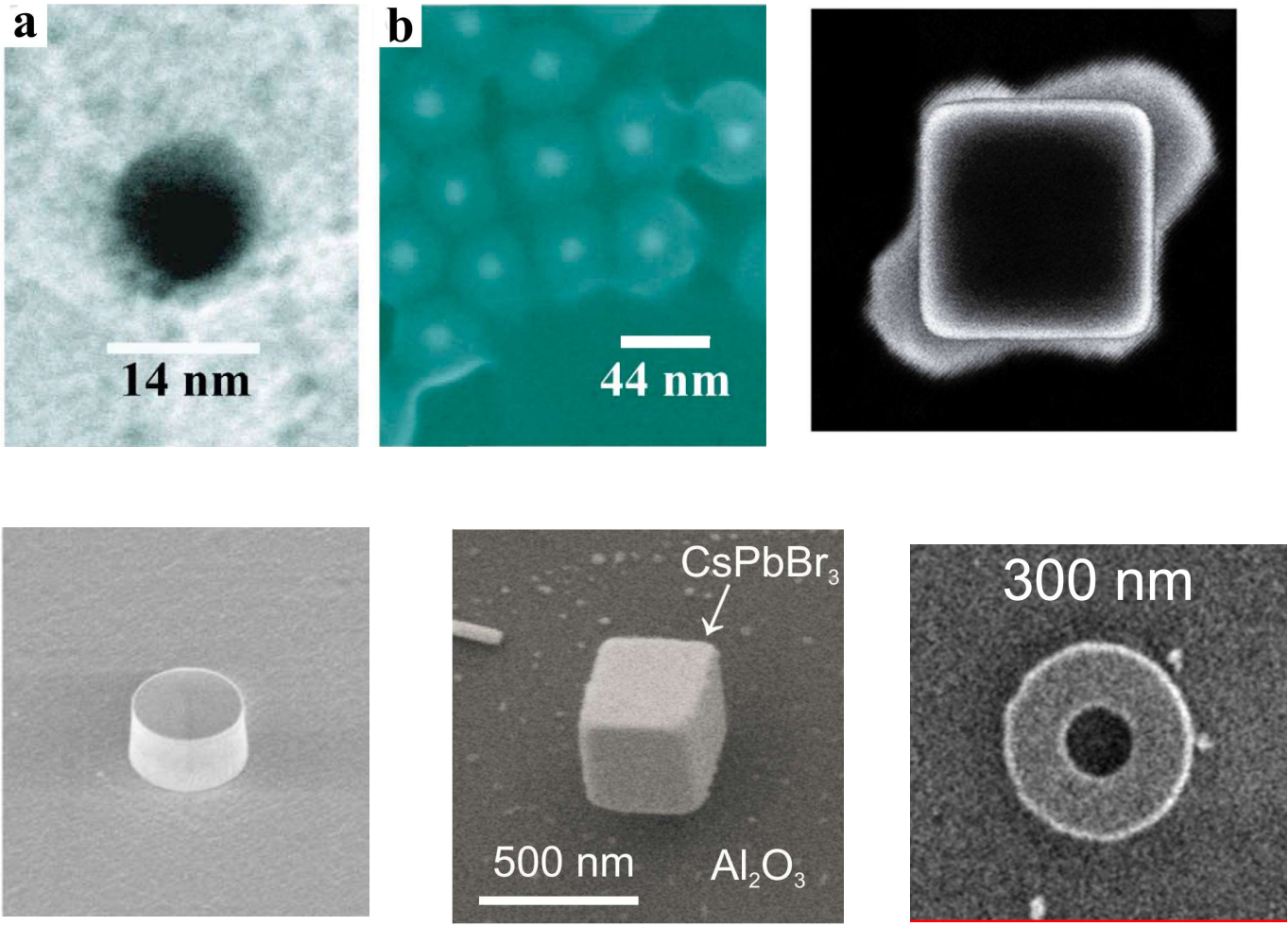

Fig.1. a) The transmission electron microscope image of a gold spherical nanoresonator with a diameter of 14 nm [2]; b) the scanning electron microscope (SEM) image of a spaser based on the resonator shown in a)[2]; c) the SEM image of a gold single-crystal nanoresonator with an edge of 100 nm with opposite edges covered by a polymer doped with CdSe/CdS/Zn (core/shell/shell) colloidal quantum dots [3]; d) the SEM image of a cylindrical AlGaAs nanoresonator (refractive index $n$=3.4) 635 nm high and 930 nm in diameter on a quartz substrate [20, 21]; e) the SEM image of a $CsPbBr_3$ cubic nanoresonator with an edge of 310 nm on a sapphire substrate [22]; f) the SEM image of a Si ring resonator with outer and inner diameters of 800 and 300 nm, respectively, and the thickness of 80 nm [23].

Therefore, the purpose of this review is to analyze and generalize the most modern approaches to the description and development of open nanoresonators, that is, resonators of the size smaller than the wavelength of radiation in vacuum. A large flow of work in this direction leads to the fact that their character is mainly phenomenological and/or purely computational. Therefore, one of the important objectives of the review is to present the situation from a single, well-defined point of view. In doing so, we restrict ourselves to the case of resonators of simple shapes with high symmetry, allowing a more or less accurate description of their resonant properties. The properties of asymmetric resonators are also very interesting (see, for example, [42]), but they critically (even chaotically) depend on small shape variations, and their study deserves a separate review.

Section 2 of the review will present general approaches to the description of optical nanoresonators. Here, various definitions and methods for finding their eigenmodes and eigenvalues will be analyzed. Particular attention will be paid to works where alternative methods for describing eigenmodes are developed that do not have the disadvantages inherent to the standard description. This section will also analyze the underlying software for finding eigenmodes and eigenvalues. Section 3 will present works studying eigen oscillations in plasmonic nanoresonators of various shapes (spheres, spheroids, ellipsoids, Platonic solids, clusters of nanoparticles). In Section 4, we will consider investigations of eigen oscillations in dielectric nanoresonators of various shapes (spheres, spheroids, cylinders). Section 5 will analyze studies of eigen oscillations in nanoresonators made from chiral and hyperbolic metamaterials, as well as from metamaterials with a negative refractive index. Section 6 of the review will present examples of the use of nanoresonators for nanolasers, biosensors and nonlinear optical devices, optical and quantum computers. Throughout the review, it is assumed that the dependence of fields on time has the form $\exp(-i\omega t)$.

## 2. General Approaches to Describing Optical Nanocavities Properties

### 2.1. Closed resonator eigenmodes

Description of the properties of electromagnetic waves in resonators is a classical problem of mathematics and the theory of electromagnetic waves (see, for example, [43-45]). In classical works, the main attention is paid to closed resonators, that is, resonators without radiation losses.

Natural oscillations of such resonators are described by the Maxwell equations with the condition of perfect conductivity of walls:

$$\nabla \times \nabla \times \mathbf{E}_n = k_{0,n}^2 \varepsilon \mathbf{E}_n$$
$$\mathbf{E}_{n,\tan}\big|_S = 0 \tag{1}$$

For resonators without internal losses, $\text{Im}\,\varepsilon = 0$, the eigenfrequencies of system (1) $\omega_n = ck_{0,n}$ are real numbers, and the modes for different frequencies are orthogonal. This makes it possible to solve all practically interesting problems concerning the excitation of such resonators and the interaction of waves in them. Losses in the walls of waveguides can be considered using perturbation theory [43]. The theory of closed resonators is the basis for considering any resonators

2.2. Open Nanoresonator Eigenmodes

In addition to closed resonators, both in microwave technology and in modern applications of nano-optics, open nanoresonators, where natural oscillations lead to the emission of energy and the associated damping of oscillations, are of great importance [46].

*2.2.1 What are open nanoresonators modes?*

The concept of a "resonator mode" seems to be well known, and this is true for closed resonators that are used in microwave technology. However, when applied to optical nanoresonators based on nanoparticles, the concept of "resonator mode" becomes far from trivial, since they are open systems [46] and oscillations in the resonator volume interact with a continuum of free space modes. So, the definition of modes in open resonators as solutions of Maxwell's equations without sources becomes incomplete, since for open resonators it is also necessary to specify the character of the behavior of fields at infinity, that can be very different both formally and when considering specific problems. Setting certain conditions at infinity imposes restrictions on the choice of spectral parameters that can be eigenvalues of the resonator modes.

In any case, the main characteristics of the resonators are the resonant frequencies $\omega_n$ and the quality factors $Q_n$, according to the definition [43], equal to the ratio of the energy stored in the *n*-th mode of the resonator $W_{n,stored}$ to the power $P_{n,rad}$ radiated by the same mode:

$$Q_n = \omega_n W_{n,stored} / P_{n,rad} \qquad (2)$$

This definition assumes that the eigenmodes have already been found in one way or another. However, in practice, in the case of open resonators, it is often difficult to find natural oscillations directly, and in this case the concept of a generalized quality factor turns out to be extremely useful, being valid at any frequencies [47-49] in contrast to (2):

$$Q(\omega) = \omega W_{stored}(\omega) / P_{rad}(\omega) \qquad (3)$$

In (3), the stored energy $W_{stored}(\omega)$ and the radiated power $P_{rad}(\omega)$ can be found by solving the scattering problem at an arbitrary frequency. At frequencies corresponding to the frequencies of natural oscillations, this expression naturally coincides with the usual definition (1). However, the concept of a generalized $Q$-factor turns out to be very useful in non-trivial cases when the stored energy does not have resonant properties and radiation losses have minima (see sections 4.1.4, 4.2.2, 4.3.3, 5.1). Calculation of the energy stored in the resonator, $W_{stored}$, for open resonators can be carried out in various ways [50]. In this review by the stored energy we mean, as usual [43], the positive energy that is concentrated inside the resonator.

*2.2.2. Frequency as mode eigenvalue*

In the case of open resonators - and this is the case that is the subject of this review - it is usually assumed (as in the theory of closed resonators) that the eigenvalue is the oscillation frequency and the modes are defined as solutions of the Maxwell equation

$$\begin{aligned} \nabla \times \nabla \times \mathbf{E}_n &= k_{0,n}^2 \varepsilon \mathbf{E}_n - \text{inside nanoresonator} \\ \nabla \times \nabla \times \mathbf{E}_n &= k_{0,n}^2 \mathbf{E}_n - \text{outside nanoessonator} \end{aligned} \qquad (4)$$

with Sommerfeld radiation conditions at infinity [51]:

$$\mathbf{E}(r, \theta, \varphi) \xrightarrow[r \to \infty]{} \frac{e^{ik_{0,n}r}}{r} \mathbf{F}(\theta, \varphi) \qquad (5)$$

In(4) $\varepsilon$ stands for resonator permittivity.

However, such a formulation of the problem is contradictory, since due to radiation and the energy conservation law it follows that the eigenfrequencies $\Omega_n$ are complex numbers, $\Omega_n = \omega_n - i\Gamma_n/2$, leading to unlimited increase of the mode fields at infinity in its turn.

$$E_n(\mathbf{r}) \sim \frac{\exp(i\Omega_n r/c)}{r} = \frac{\exp(i\omega_n r/c)\exp(\Gamma_n r/2c)}{r} \xrightarrow[r \to \infty]{} \infty \qquad (6)$$

For the first time, L.A. Weinshtein drew attention to this fact [46]: "This increase is an inevitable consequence of the exponential damping of the field as $t \to \infty$. Indeed, since at $R \to \infty$ the field of each natural oscillation has the character of a spherical wave going to infinity at the speed of light, then, for example, at $t=0$, the field at a large distance $R$ is due to a wave emitted by a sphere at $t=-R/c$ when the oscillation amplitude in it was much greater than at $t=0$, due to which this field is exponentially large. The above considerations are very general and allow us to assert that the exponential increase in the field of eigen oscillations at $R \to \infty$ ..... takes place for all open resonators (three-dimensional)."

Eigenmodes of open systems are also known as decaying states, leaky modes, quasi-modes, or quasi-normal modes.

The main problem in the description and computer simulation of quasi-modes or modes of open resonators is to determine how they can be normalized. Since these modes increase at infinity, the usual normalization procedures do not work, and non-standard solutions are needed.

A number of approaches to the normalization of quasi-modes have been proposed in [46, 52-59]. In [56], the eigenfunctions of spherical resonators, that are the solutions of the Maxwell equations without sources:

$$\nabla \times \nabla \times \mathbf{E}_n - k_{0,n}^2 \varepsilon(|\mathbf{r}|) \mathbf{E}_n = 0 \tag{7}$$

are proposed to be normalized using the relation

$$\int_V d\mathbf{r}\, \varepsilon(|\mathbf{r}|) \mathbf{E}_n^2(\mathbf{r}) + \frac{1}{2k_{0,n}^2} \int_S dS \left( \mathbf{E}_n \cdot \frac{\partial \mathbf{E}_n}{\partial r} + r \mathbf{E}_n \cdot \frac{\partial^2 \mathbf{E}_n}{\partial r^2} - r \left( \frac{\partial \mathbf{E}_n}{\partial r} \right)^2 \right) = 1 \tag{8}$$

where the integration is carried out over the spherical resonator volume $V$ and over the surface of the large sphere $S$ surrounding the volume $V$. In (8), the volume and surface integrals, taken separately, diverge, but their sum remains finite.

To find the eigenfrequencies and $Q$-factors of nanoresonators, one can use the method of integral equations (see, for example, [60-63]), where, apparently, there are no problems with field divergences at infinity, that arise when solving partial differential equations (4). However, the low prevalence of this method indicates the complexity of its application.

### 2.2.3. Permittivity as mode eigenvalue

More reliable and unambiguous results are obtained if we use not the frequency (as is usually done), but the permittivity $\varepsilon_n$ of the nanoresonator [64,65, 1] as an eigenvalue.

In this case, the eigenfunctions satisfy the Maxwell equations

$$\nabla \times \nabla \times \mathbf{E}_n = k_0^2 \varepsilon_n \mathbf{E}_n - \text{inside resonator}$$
$$\nabla \times \nabla \times \mathbf{E}_n = k_0^2 \mathbf{E}_n - \text{outside resonator} \quad (9)$$

and the Sommerfeld conditions (5) at infinity.

In (9), $\omega = k_0 c$ is the real frequency, and the permittivity eigenvalue $\varepsilon_n$ is a complex number. It can be shown [64,65,1], that its imaginary part is negative, $\text{Im}\,\varepsilon_n < 0$, and this can be interpreted as an amplifying medium that compensates radiation losses. It is this circumstance that causes the fields of eigenfunctions $\mathbf{E}_n$ to decrease at infinity. In addition, the smallest eigenvalues $\varepsilon_n$ have a negative real part, making this method especially useful for describing the properties of plasmonic nanoresonators. We emphasize once again that in system (9) the actual material properties of the resonator do not appear.

The expansion of solutions of Maxwell's equations in terms of eigenmodes for an arbitrary excitation field $\mathbf{E}^0$ in the framework of the "$\varepsilon$ - method" has the simple form:

$$\mathbf{E} = \mathbf{E}^0 + \sum_n \mathbf{E}_n \frac{(\varepsilon(\omega)-1)}{(\varepsilon_n - \varepsilon(\omega))} \frac{\int_V \mathbf{E}_n \mathbf{E}^0 dV}{\int_V \mathbf{E}_n^2 dV} \quad (10)$$

where $\varepsilon(\omega)$ describes the frequency dependence of the permittivity of the resonator's specific material, and the integration is carried out over the volume of the resonator $V$.

It follows from the form of this solution that $\varepsilon(\omega)$ enters into (10) in a rather simple way, making it easy to carry out calculations for nanoresonators of the same shape, but made of different materials. This is very important when optimizing certain nanodevices.

The most important feature of solution (10) is the presence of the resonance factor $(\varepsilon_n - \varepsilon(\omega))$ in the denominator. At frequencies $\omega_n$ such that $\varepsilon_n - \varepsilon(\omega_n) \approx 0$, the resonance occurs in the system, and only one term becomes significant in the solution:

$$\mathbf{E} \approx \mathbf{E}^0 + \mathbf{E}_n \frac{(\varepsilon(\omega)-1)}{(\varepsilon_n - \varepsilon(\omega))} \frac{\int_V \mathbf{E}_n \mathbf{E}^0 dV}{\int_V \mathbf{E}_n^2 dV} \quad (11)$$

In this case, one can speak of the excitation at a frequency $\omega_n$ of a localized mode with a spatial structure described by the mode field $\mathbf{E}_n$, that does not depend at all on a particular material!

The width of the resonance as a function of frequency essentially depends on the imaginary parts of $\varepsilon_n$ and $\varepsilon(\omega)$. It is very important to note that the imaginary parts of $\varepsilon_n$ and $\varepsilon(\omega)$ have different signs always, and therefore the resonance width of (11) is determined by the sum of their moduli. As will be shown below, in the case of nanoresonators, the imaginary part of $\varepsilon_n'' \sim (ka)^3$ becomes very small. In this case, the width of the resonance is determined by the losses inside the nanoparticles mainly, that is, by the imaginary part of $\varepsilon(\omega)$. In those cases where the imaginary part is small, it becomes necessary to take the imaginary part of $\varepsilon_n$ into account, and a number of interesting effects arise in this case [66].

Thus, within the framework of the "$\varepsilon$-approach", plasmon oscillations arise naturally as a result of solving of the spectral problem where the eigenvalue is the permittivity, not the frequency. Another important feature of the described approach is that the properties of plasmons depend actually only on the shape of the particle, and not on the specific material of the resonator. Both these factors make the "ε-approach" extremely useful in a wide variety of applications and problems. Moreover, one can say that the "$\varepsilon$-approach" is a constructive definition of localized plasmons.

*2.2.4. Perfect nonradiating modes*

Usually, the modes of open resonators are found by solving the homogeneous Maxwell equations with the Sommerfeld radiation conditions at infinity (5) and, therefore, such modes are *fundamentally* related to radiation losses. Moreover, such modes increase unlimitedly at infinity, requiring the development of very complex artificial approaches to describe them (see, for example, [46, 52-59]).

However, the finding all modes is a non-trivial task, and the quasi-normal modes studied in recent works [18-23, 27-38] do not exhaust all solutions of the sourceless Maxwell equations in open nanoresonators. Quite recently in [39-41], a fundamentally new class of eigenoscillations in nanoresonators was found - perfect nonradiating modes that are solutions of sourceless Maxwell's equations and fundamentally do not contain waves that carry energy away!

More specifically, in [40, 41], it was proposed to search for an electromagnetic field outside the nanoresonator in the form of a superposition of solutions of Maxwell's equations that are nonsingular in an unbounded free space. This approach is fundamentally different from the usual approach assuming that the functions describing the fields outside the resonator can have singularities upon analytic continuation to the region inside the resonator. For example, in the expansion of any field component $E(r,\theta,\varphi,\omega)$ outside the resonator in terms of spherical harmonics, $Y_n^m(\theta,\varphi)$, in accordance with the Sommerfeld radiation condition, the spherical Hankel functions $h_n^{(1)}(k_0 r)$, which are singular at $r=0$ (inside the resonator), are usually used:

$$E(r,\theta,\varphi,\omega) \sim \sum a_{nm} h_n^{(1)}(k_0 r) Y_n^m(\theta,\varphi) \qquad (12).$$

In [40] to find the fields of nonradiating modes outside the resonator, it was proposed to use only functions that are nonsingular inside the resonator, for example, the spherical Bessel functions, $j_n(kr)$, so that instead of (12), the asymptotics of the fields have the form:

$$E(r,\theta,\varphi,\omega) \sim \sum a_{nm} j_n(k_0 r) Y_n^m(\theta,\varphi) \qquad (13)$$

Obviously, if the solution of (4) with the asymptote (13) exists, then in principle it will not have a flux of energy and radiation at infinity!

Since (13) has no singularities in the entire space, the initial problem of finding fields in infinite space can be reduced to the problem of finding fields only inside the nanoresonator. As a result, the system of equations that determines nonradiating modes in a nanoparticle with the permittivity $\varepsilon$, can be written as a system of two equations for two auxiliary fields $\mathbf{E}_1$, $\mathbf{E}_2$ inside the region $V$, determining the resonator:

$$\begin{aligned}\nabla \times \mathbf{E}_1 = ik_0 \mathbf{H}_1; \nabla \times \mathbf{H}_1 = -ik_0 \varepsilon \mathbf{E}_1, \mathbf{r} \in V \\ \nabla \times \mathbf{E}_2 = ik_0 \mathbf{H}_2; \nabla \times \mathbf{H}_2 = -ik_0 \mathbf{E}_2, \mathbf{r} \in V\end{aligned} \qquad (14)$$

connected only through the boundary conditions for the continuity of the tangential components of the electric and magnetic fields

$$\mathbf{E}_{t,1} = \mathbf{E}_{t,2}; \mathbf{H}_{t,1} = \mathbf{H}_{t,2} \qquad (15)$$

on the surface of the resonator.

At some real values of frequency or permittivity, the system (14), (15) becomes consistent, meaning the appearance of "perfect" nonradiating modes. The physical fields inside the resonator are equal to $\mathbf{E}_1, \mathbf{H}_1$, while the physical fields outside the particle are determined by the analytical continuation of the auxiliary solution $\mathbf{E}_2, \mathbf{H}_2$ into outside region.

The modes found in this way form an orthogonal system and have no analogues. In particular, they differ from the so-called anapole current distributions [19,31-36], in that, unlike the latter, their fields outside the particle are non-zero and have well-defined expansions in spherical harmonics (13).These modes also differ from the "bound states in a continuum" mode, as they do not decay exponentially. The new modes are closest to the strange (merkwiirdige) Neumann-Wigner modes [67], but unlike the latter, the nanoresonator potential (permittivity) differs from the free space value only in a finite region of space, which fundamentally distinguishes perfect nonradiating modes from strange Neumann-Wigner modes [67], where the potential is nonzero in the entire space. The complete absence of radiation makes it possible to call these modes perfect nonradiating modes.

The system of equations (14)(15) is very complicated, and there is no rigorous mathematical theory for it in the general case. Nevertheless, in [40,41], it was possible to find conditions for the existence of perfect nonradiative modes for arbitrary spheroids, hyperspheroids, and elliptical cylinders , describing well almost all nanoparticle shapes of interest for applications. Perfect nonradiating modes are not abstract solutions, they are of great importance for finding the conditions when the scattered power becomes minimal or even zero, while the energy stored in the nanoresonator remains finite. This leads to an unlimited increase in the *Q*-factor of these modes (see Sections 4.1.2, 4.2.2, 4.3.3).

2.3. Anapole Current Distributions and Eigenmodes

The concept of anapole was introduced by Zel'dovich [68] to designate a current with electromagnetic fields equal to zero where this current is absent. Anapole is the simplest representative of the family of Cartesian toroidal (anapole) multipoles, necessary (along with Cartesian electric and magnetic multipoles) for a complete description of the field of arbitrary current sources. An illustrative model of a toroidal anapole can be a torus-shaped solenoid with the current flowing through its winding. A change in the anapole (toroidal) moment with time leads, in the general case, to the emission of electromagnetic radiation by the system waves. However, there are such distributions of charge densities and currents, $\rho(\mathbf{r}), \mathbf{J}(\mathbf{r})$, when the fields of Cartesian electric multipoles

$$Q_{l,m} \sim \int Y_{lm}^* \frac{\partial}{\partial r}\left[rj_l(kr)\right]\rho(\mathbf{r})d^3\mathbf{r} \qquad (16)$$

and Cartesian toroidal multipoles

$$T_{l,m} \sim \int Y_{lm}^* j_l(kr)\left[\mathbf{r}\cdot\mathbf{J}(\mathbf{r})\right]d^3\mathbf{r} \qquad (17)$$

completely cancel each other out! Sometimes such current distributions are called anapole states or even anapole modes (see Section 4.1.3). Such definitions, of course, are not correct, since modes, by definition, are solutions of the sourceless Maxwell equations.

2.4. Symmetry properties of modes in optical nanoresonators [69,70]

The spatial structure of the eigenmodes of the resonators is rigidly related to the symmetry of the shape of the resonators. In [69,70] an algorithm was developed for classifying eigenmodes in resonators of the simplest shapes depending on their symmetry group. For each mode type $\mathbf{E}_n$, its vector multipole content is found:

$$\mathbf{E}_n(\mathbf{r}) = \sum a_n \mathbf{M}_{nmp}(\mathbf{r}) + \sum b_n \mathbf{N}_{nmp}(\mathbf{r}) \qquad (18)$$

where **M** and **N** are the vector spherical harmonics of the magnetic and electric types, respectively [51]. Relation (18) creates a bridge between modal and multipole descriptions.

The authors of [69] claim that their approach can be used to design, predict, and explain the scattering phenomena and optical properties of nanoresonators based only on their symmetry without the need for numerical simulations. However, despite the useful qualitative picture of the multipole composition of modes of a certain symmetry, without numerical analysis, it is apparently impossible to say whether the multipoles in (18) are suppressed or not for specific resonator shapes (see also Section 4.3.2).

2.5. Perturbation theory methods for describing the properties of nanoresonators
 2.5.1. Rayleigh method (quasi-statics)

As early as 1897, Rayleigh showed [71], that to describe the scattering of light by nanoparticles one can often use the perturbation theory with respect to a small parameter

$$ka = \omega a / c = 2\pi a / \lambda \ll 1, \qquad (19),$$

where $a$ is the characteristic size of a nanoparticle and $\lambda$ is the wavelength of radiation in the surrounding space. In [72,73], this method was further developed. As applied to nanoresonators, this approach can be formulated as follows [1]. Within the framework of this approach, all fields are searched in the form of series in powers of $k$:

$$\mathbf{E} = \mathbf{E}^{(0)} + k\mathbf{E}^{(1)} + k^2\mathbf{E}^{(2)} + ...; \quad \mathbf{H} = \mathbf{H}^{(0)} + k\mathbf{H}^{(1)} + k^2\mathbf{H}^{(2)} + ... \qquad (20)$$

while the permittivity $\varepsilon(\omega)$ is not expanded in powers of $k$ (frequency).

Further, by substituting such series into Maxwell's equations and equating to zero the terms at the same powers of $k$, the system of Maxwell's equations can be reduced to a set of potential theory problems. In particular, the eigenvalues $\varepsilon_n$ and the eigenfunctions of the "$\varepsilon$ method" (see section 2.2.3) can be found in the first approximation by solving the equations

$$\nabla(\varepsilon(\mathbf{r})\mathbf{E}_n) = 0; \quad \nabla \times \mathbf{E}_n = 0, \qquad (21)$$

which, using substitution, $\mathbf{E}_n = -\nabla \varphi_n$ can be reduced to the solution of the Laplace equations

$$\Delta \varphi_n^{in} = 0; \quad \Delta \varphi_n^{out} = 0; \quad \varepsilon_n \left.\frac{\partial \varphi_n^{in}}{\partial \mathbf{n}}\right|_S = \bar{\varepsilon} \left.\frac{\partial \varphi_n^{out}}{\partial \mathbf{n}}\right|_S \qquad (22)$$

In (22), $\bar{\varepsilon}$ is the permittivity of the space surrounding the nanoresonator, $\varphi_n^{in}, \varphi_n^{out}$ stand for the potentials of the eigenfunctions inside and outside the particle, respectively, and $\left.\frac{\partial \varphi_n}{\partial \mathbf{n}}\right|_S$ denotes the normal derivative at the boundary of the particle. The last equation in (22) ensures that the normal components of the induction or the tangential components of the magnetic field are continuous. In this case, the solution of Maxwell's equations with given excitation fields (10) remains the same. Note that in this case to describe plasmon oscillations, there is no need to find magnetic fields at all.

The quasi-static description of plasmon resonances (22) in nanoparticles is much simpler than the complete system of Maxwell's equations, because instead of the Helmholtz equations, one has to solve the Laplace and Poisson equations. The solution of the Laplace and Poisson equations can be found for nanoresonators of various shapes.

The most important feature of the quasi-static description (22) is that it allows dealing only with plasmon oscillations. Other particle modes (whispering gallery modes, etc.) do not appear in this description and do not make it difficult to obtain and interpret the results.

From a mathematical point of view, the quasi-static "$\varepsilon$-approach" differs from the full "$\varepsilon$-approach" in that in the former, the resonant values of the permittivity $\varepsilon_n$ are negative real numbers. The absence of an imaginary part in the resonant values of the permittivity $\varepsilon_n$ is due to the fact that there is no radiation in the quasi-static approximation, while the imaginary parts of $\varepsilon_n$ arise precisely due to radiation losses.

Based on the foregoing, we conclude that the quasi-static "$\varepsilon$-approach" describes well exactly the plasmon part of the spectrum and can be effectively used to determine the properties of localized plasmons.

Naturally, the substitution of $\varepsilon_n$ and mode distributions $\mathbf{E}_n = -\nabla \varphi_n$ found in the quasi-static approximation into (11) makes it possible to find fields only in the near zone. Nevertheless, quasi-static approach allows to find not only the natural frequencies $\omega_n$ of eigen oscillations (by solving the equation $\varepsilon(\omega_n) = \varepsilon_n$), but also their $Q$-factors (2), where the expressions for the energy stored in the resonator and the radiated power have the form:

$$W_{n,stored} = \frac{d(\omega_n \varepsilon(\omega_n))}{d\omega_n} \int_V dV |\mathbf{E}_n|^2; \quad P_{n,rad} = \frac{\omega_n k_n^3}{3} |\mathbf{d}_n|^2; \quad \mathbf{d}_n = \frac{\varepsilon(\omega_n)-1}{4\pi} \int_V dV \mathbf{E}_n \quad (23)$$

where $\mathbf{d}_n$ is the electric dipole moment of the $n$-th mode of the nanoresonator.

*2.5.2. Perturbation theory for large permittivity*

The Rayleigh-Stevenson method (see the previous section) can also be applied to dielectric resonators with a large positive permittivity, when the dimensions $L$ are small with respect to the wavelength $\lambda_{diel}$ inside the body , $L \ll \lambda_{diel} = \lambda_0 / N$, where $N$ is the refractive index of the nanoresonator, $N = \sqrt{\varepsilon} \gg 1$. This limitation obviously excludes the study of nanocavities with a high permittivity, where the resonator size is smaller than the wavelength in vacuum $\lambda_0$, but larger than the wavelength in the resonator material $\lambda_0 \gg L \geq \lambda_{diel} = \lambda_0 / N$.

As the refractive index $N$ tends to infinity, it makes sense to focus on a specific resonant mode corresponding to a finite wave number $k$ in the dielectric to understand what happens with the fields as $N \to \infty$ [74]. During this passage to the limit, $kL$ approaches an asymptotic value that is finite and non-zero. This value is a characteristic of the mode. The wave number $k_0$ in vacuum during such a passage $(k_0 = 2\pi / \lambda_0 = k / N = \omega / c_0)$ approaches zero along with the frequency, and the wavelength $\lambda_0$ tends to infinity.

In [74], it was proposed to search for the modes of such nanoresonators by expanding the fields into series:

$$\mathbf{H} = \mathbf{H}_0 + \mathbf{H}_2 / N^2 + ...; \quad \mathbf{E} = \mathbf{E}_1 / N + \mathbf{E}_3 / N^3 + ... \qquad (24)$$

As a result, in the zeroth approximation, the natural oscillation frequencies $\omega_m = ck_m / N$ and their spatial structure can be found from the system of equations:

$$\begin{aligned} \nabla \times \nabla \times \mathbf{H}_m &= k_m^2 \mathbf{H}_m - \text{inside resonator} \\ \nabla \times \mathbf{H}_m &= 0 - \text{outside resonator} \\ \nabla \cdot \mathbf{H}_m &= 0 - \text{everywhere} \end{aligned} \qquad (25)$$

In the particular case of axisymmetric nanoresonators, there can be a subset of modes - the so-called confined modes with the magnetic fields equal to zero outside the resonator in the limit $\varepsilon \to \infty$ and satisfying the equations:

$$\begin{aligned} \nabla \times \nabla \times \mathbf{H}_m &= k_m^2 \mathbf{H}_m - \text{inside the resonator} \\ \mathbf{H}_m &= 0 \quad - \quad \text{at the resonator boundary} \end{aligned} \qquad (26)$$

The electric fields have a higher order of smallness and can be found from the relation:

$$\mathbf{E}_m = \frac{i}{k_m N} \nabla \times \mathbf{H}_m \qquad (27)$$

In this case, despite the tendency of the electric field to zero, the energy of the electric field in the resonator remains finite.

Van Bladel's approach [74] is extremely important, not only because of the possibility of constructing a good perturbation theory, but also because it allows one to understand the general properties of arbitrary shaped dielectric nanoresonators. In particular, using this approach, van Bladel [74] showed that for resonators with a very high permittivity, the radiation quality factor $Q_{rad}$ depends on $\varepsilon$ as

$$Q_{rad} \sim \varepsilon^P \qquad (28)$$

where $P = 1.5$ for modes radiating as a magnetic dipole (nonconfined modes, see (25) and $P = 2.5$ for modes radiating as an electric dipole (confined modes (26)), as well as for modes radiating like a magnetic quadrupole.

Equation (28) and the $P$ values given above are quite general and are valid regardless of the shape of the resonator. However, they give good accuracy only for large values of $\varepsilon$. For example, for confined TM modes of axisymmetric nanoresonators with $\varepsilon = 80$ from (28) we get $Q=57000$!

In [75], based on (28), more accurate approximation formulas were found for the quality factors of dielectric nanocylinders with different aspect ratios.

### 2.5.3. Generalized Brillouin-Wigner Perturbation Theory

Above, the perturbation theories were considered based on the smallness of the resonator dimensions relative to the wavelength or on the smallness of $1/\varepsilon$ for the material of the resonator. The Wigner-Brillouin perturbation theory [28, 69, 76, 77] is based on the smallness of shape variations of resonators with known set of eigenfunctions and eigenvalues. As a rule, spherical resonators are chosen as unperturbed resonators. For a sphere, a complete set of eigenfunctions is known - these are vector spherical harmonics with normalization (8). The expansion of any mode of a resonator with permittivity inhomogeneities $\Delta\varepsilon(\mathbf{r})$

$$\nabla \times \nabla \times \mathbf{E}_n = \left(\varepsilon^* + \Delta\varepsilon(\mathbf{r})\right)\frac{\Omega_n^2}{c^2}\mathbf{E}_n \qquad (29)$$

can be represented as

$$\mathbf{E}_n(\mathbf{r}) = \sum_s C_{ns}\mathbf{Y}_s(\mathbf{r}) \qquad (30)$$

where $\Omega_n$ is the complex eigenfrequency of the mode $\mathbf{E}_n$, $\mathbf{Y}_s(\mathbf{r})$ is the eigenfunction of a sphere with permittivity $\varepsilon^*$, and complex eigenfrequency $\omega_s$, and $s$ is the composite index that includes polarization and orbital, azimuthal and radial quantum numbers.

From the generalized Wigner-Brillouin perturbation theory, the dispersion equation can be obtained:

$$\frac{1}{\omega_{s'}}\sum_s C_{ns}\left[\delta_{ss'} + V_{ss'}\right] = \frac{1}{\Omega_n}C_{ns'} \qquad (31)$$

where the overlap integral

$$V_{ss'} = \frac{1}{2\varepsilon^*} \int dV \Delta\varepsilon(\mathbf{r}) \mathbf{Y}_s \cdot \mathbf{Y}_{s'} \quad (32)$$

describes the interaction between different unperturbed modes of a spherical resonator due to its small deformation.

The inhomogeneity of the permittivity, $\Delta\varepsilon(\mathbf{r})$, can be chosen so that, in fact, (29) will describe a resonator different from spherical one. It is important that this resonator should be inside the initial spherical resonator, otherwise the convergence will be poor. For example, if we choose $\Delta\varepsilon = 1 - \varepsilon^*$ in the region between the cylinder inscribed in the original sphere and the surface of the sphere, then (31) will describe the natural oscillations of the cylinder (see Fig. 2). Applications of this method will be discussed in Section 4.

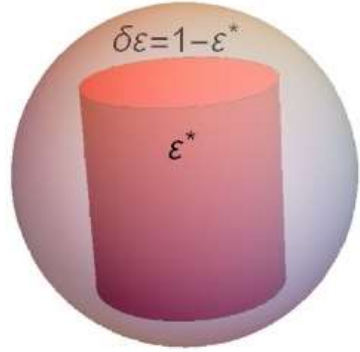

Fig.2. An illustration of the application of Brillouin-Wigner perturbation theory by the example of a cylinder inside a spherical region. A perturbation $\Delta\varepsilon = 1 - \varepsilon^*$ outside the cylinder is introduced into a sphere with permittivity $\varepsilon^*$. As a result, the perturbed resonator has the shape of a cylinder with permittivity $\varepsilon^*$.

2.6. Numerical methods for describing the properties of optical nanoresonators

Usual quasi-normal modes, their eigenfrequencies and $Q$-factors are relatively easy to find by calculating spectra of light scattering by nanoparticles making use of the commercial packages Comsol [78] or CST Studio Suite [79]. Also, these packages allow one to set a specific spatial structure of incident beams for excitation and study of modes that cannot be detected under a plane wave illumination.

The same packages can also be used conveniently to find expansion coefficients in various methods of perturbation theory. A more detailed analysis of usual modes of open resonators using the same packages can be found, for example, using the numerical approach proposed in [80].

To analyze natural frequencies, one can also use the numerical solution of surface or volume integral equations. These methods are highly efficient because they have smaller dimensions. However, these algorithms are complex and must be implemented manually (see, for example, [61]).

## 3. Modes in Plasmonic Nanoresonators

The uniqueness of plasmonic nanoresonators lies in the fact that metal nanoparticles have natural oscillations frequencies in the optical range, from ultraviolet to infrared ranges. On the other hand, there are many methods for creating plasmonic nanoparticles, including colloid chemistry and electron nanolithography. Finally, the properties of plasmonic nanoresonators can be understood in the first approximation using the quasi-static approximation (see Section 2.5.1). All this led to the rapid development of research on the resonance properties of plasmonic nanoparticles [1].

### 3.1. Optical Properties of Spherical Plasmonic Resonators

The spherical geometry is unique and allows one to describe in detail all the optical properties of spherical nanoresonators, which are completely determined by the poles of the reflection coefficients of spherical waves of different polarization incident upon the spherical particle of an arbitrary size and composition [51,82, 1]:

$$q_n = \frac{\varepsilon_1 \frac{d}{dz_2}[z_2 j_n(z_2)] j_n(z_1) - \varepsilon_2 \frac{d}{dz_1}[z_1 j_n(z_1)] j_n(z_2)}{\varepsilon_1 \frac{d}{dz_2}[z_2 h_n^{(1)}(z_2)] j_n(z_1) - \varepsilon_2 \frac{d}{dz_1}[z_1 j_n(z_1)] h_n^{(1)}(z_2)} \text{TM polarization}, \quad (33)$$

$$p_n = \frac{\mu_1 \frac{d}{dz_2}[z_2 j_n(z_2)] j_n(z_1) - \mu_2 \frac{d}{dz_1}[z_1 j_n(z_1)] j_n(z_2)}{\mu_1 \frac{d}{dz_2}[z_2 h_n^{(1)}(z_2)] j_n(z_1) - \mu_2 \frac{d}{dz_1}[z_1 j_n(z_1)] h_n^{(1)}(z_2)} \text{TE polarization} \quad (34)$$

Here, $z_{1,2} = \sqrt{\varepsilon_{1,2} \mu_{1,2}} k_0 a = k_{1,2} a$, $a$ is the resonator radius, and $\varepsilon_{1,2}, \mu_{1,2}$ stand for permittivity and permeability of the sphere (1) and the surrounding space (2), respectively.

For plasmonic nonmagnetic nanoresonators in vacuum, the dispersion equations for the frequencies of dipole, quadrupole, and octupole resonances can be found from (34) by expanding denominators over the small size parameter $k_0 a$ [1]:

$$\varepsilon_{res,1}(\omega) + 2 + \frac{12}{5}(k_0 a)^2 + 2i(k_0 a)^3 + .. = 0$$

$$\varepsilon_{res,2}(\omega) + \frac{3}{2} + \frac{5}{14}(k_0 a)^2 + \frac{65}{392}(k_0 a)^4 + \frac{i}{12}(k_0 a)^5 + .. = 0 \qquad (35)$$

$$\varepsilon_{res,3}(\omega) + \frac{4}{3} + \frac{56}{405}(k_0 a)^2 + \frac{11788}{601425}(k_0 a)^4 + \frac{469672}{95954625}(k_0 a)^6 + \frac{4i}{2025}(k_0 a)^7 + .. = 0$$

where, as usual $k_0 = \omega/c$. It is important to note that the imaginary parts of the resonant permittivity (which are associated with to radiation losses) appear in higher and higher orders in powers of $k_0 a$ with increasing multipole order, associating with the smallness of the corresponding radiation intensities and leading to high $Q$-factors of such natural oscillations.

Plasmon oscillations under conditions (35) take place for arbitrarily small spheres and are of great importance in various applications.

Potentials of quasi-static electric fields corresponding to resonances (35) can be found from (22) and have a simple form:

$$\Phi_n^m = \begin{cases} (r/a)^n Y_n^m(\theta,\varphi), & r < a \\ (a/r)^{n+1} Y_n^m(\theta,\varphi), & r > a \end{cases} \qquad (36)$$

which is very similar to the solution of the quantum mechanical problem for the hydrogen atom.

The expressions (35),(36) together with the solution of the general scattering problem within the framework of the "$\varepsilon$–approach" (11) give a good description of all optical properties of spherical plasmonic nanoresonators.

Despite that the electric fields of natural oscillations in plasmonic nanoresonators are described by simple spherical harmonics, the energy flows have a rather complicated form even for very small nanoresonators. In particular, it was shown in [83-85] that near the resonances there are non-trivial vortices in the pattern of energy flows inside the resonator and around it. When weak dissipation is taken into account, these vortices terminate inside the sphere [86].

A very interesting regime of scattering of a converging spherical wave $h_n^{(2)}(k_0 r)$ by a plasmonic sphere was found in [87], where it was shown that, if the condition

$$\varepsilon_1 \frac{d}{dz_2}\left[z_2 h_n^{(2)}(z_2)\right] j_n(z_1) = \varepsilon_2 \frac{d}{dz_1}\left[z_1 j_n(z_1)\right] h_n^{(2)}(z_2) \tag{37}$$

is fulfilled, the reflected spherical wave proportional to $h_n^{(1)}(k_0 r)$ is completely absent and the solution of Maxwell's equations turns into a single vortex, starting at the infinity and ending at the center of the sphere, that is, the coherent perfect absorption appears. Note that condition (37) coincides with the condition of vanishing of the complex conjugate denominator of the Mie coefficient (33), and therefore the solution found in [87] can be considered as a time-reversed solution for a spaser [2].

A converging spherical wave, $h_n^{(2)}(k_0 r)$ resulting in perfect absorption, cannot be accurately realized in practice. Therefore, to realize perfect coherent absorption by a plasmonic nanoparticle, a specially selected superposition of radially polarized Bessel beams in the form

$$\mathbf{E}_{in}(\mathbf{r}) = \begin{pmatrix} E_\rho \\ E_\varphi \\ E_z \end{pmatrix} = \int_0^{k_0} dk_\rho E_0(k_\rho) \begin{pmatrix} ik_z J_1(k_\rho \rho) \\ 0 \\ k_\rho J_0(k_\rho \rho) \end{pmatrix} e^{-i\sqrt{k_0^2 - k_\rho^2}\, z} \tag{38},$$

was used in [88]. In (38), the weight function $E_0(k_\rho)$ should be chosen to achieve complete absorption. Assuming that the absorbing nanoparticle can be described in the dipole approximation, an explicit form for $E_0(k_\rho)$, was found in [88].

### 3.2. Optical Properties of Spheroidal and Ellipsoidal Plasmonic Resonators

The optical properties of spheroidal and ellipsoidal plasmonic resonators can also be found in the framework of the quasi-static approximation.

For prolate spheroids with semi-axes $a=b<c$, the dispersion equation that determines the frequencies of plasmon oscillations has the form [1]

$$\varepsilon(\omega_n^m) = \varepsilon_n^m = \frac{\left(Q_n^m(\xi_0)\right)' P_n^m(\xi_0)}{\left(P_n^m(\xi_0)\right)' Q_n^m(\xi_0)} \tag{39}$$

where the parameter $\xi_0 = c/\sqrt{c^2 - a^2}$, and $P_n^m, Q_n^m$ are the associated Legendre polynomials of the first and second kind, respectively.

Corresponding to (39), the distributions of mode potentials have the form:

$$\varphi_n^m = P_n^m(\eta) P_n^m(\xi) Q_n^m(\xi_0) \begin{bmatrix} \cos m\varphi \\ \sin m\varphi \end{bmatrix}, \xi < \xi_0$$

$$\varphi_n^m = P_n^m(\eta) P_n^m(\xi_0) Q_n^m(\xi) \begin{bmatrix} \cos m\varphi \\ \sin m\varphi \end{bmatrix}, \xi > \xi_0$$

(40)

In (40), $\eta, \xi, \varphi$ are prolate spheroidal coordinates [89]. For a plasmonic nanoresonator in the form of an oblate spheroid, these expressions are also valid after the corresponding analytical continuation.

The most common form of a plasmonic nanoresonator with known natural frequencies is a triaxial nanoellipsoid [1,90, 91]. For an arbitrary plasmonic resonator in the form of a nanoellipsoid with semiaxes $a_1 > a_2 > a_3$ along each axe of the Cartesian coordinate system $x_1, x_2, x_3$, the dispersion equation that determines the frequencies of plasmon oscillations $\omega_n^m$ has the form:

$$\varepsilon(\omega_n^m) = \varepsilon_n^m(a_1, a_2, a_3) = \frac{E_n^m(a_1; a_1, a_2, a_3) F_n'^m(a_1; a_1, a_2, a_3)}{E_n'^m(a_1; a_1, a_2, a_3) F_n^m(a_1; a_1, a_2, a_3)} \quad (41)$$

For the eigenfunctions of plasmon oscillations, respectively, we have the expressions

$$\varphi_n^m = E_n^m(\mu; a_1, a_2, a_3) \times E_n^m(\nu; a_1, a_2, a_3) \times \begin{cases} F_n^m(a_1; a_1, a_2, a_3) \times E_n^m(\rho; a_1, a_2, a_3), & \rho \leq a_1 \\ E_n^m(a_1; a_1, a_2, a_3) \times F_n^m(\rho; a_1, a_2, a_3), & \rho > a_1 \end{cases}$$

(42)

In (41) (42), $F_n^m, E_n^m$ are the external and internal Lamé functions, respectively, and $\rho, \mu, \nu$ are ellipsoidal coordinates [92], the values of $n$ are natural numbers, and $m$ varies from 1 to $2n+1$, as in spherical harmonics. Figure 3 [90] shows the dependencies of the eigenfrequencies of the lowest plasmon oscillations on aspect ratio $a_3/a_1$ in a nanoellipsoid with the permittivity described by the Drude law:

$$\varepsilon(\omega) = 1 - \omega_{pl}^2 / \omega^2 \quad (43)$$

In [90], plots of eigenfrequencies for higher-order modes are also presented.

In a triaxial ellipsoid, there is no axial symmetry, and therefore the eigenvalues degenerated in the case of a sphere split into $2n + 1$ different ones. Moreover, as the parameters of the ellipsoid change, these eigenvalues change in a non-monotonic and non-trivial way.

This is extremely important, since it enables effective control of optical processes where several frequencies take place, for example, to create bright artificial fluorophores [91].

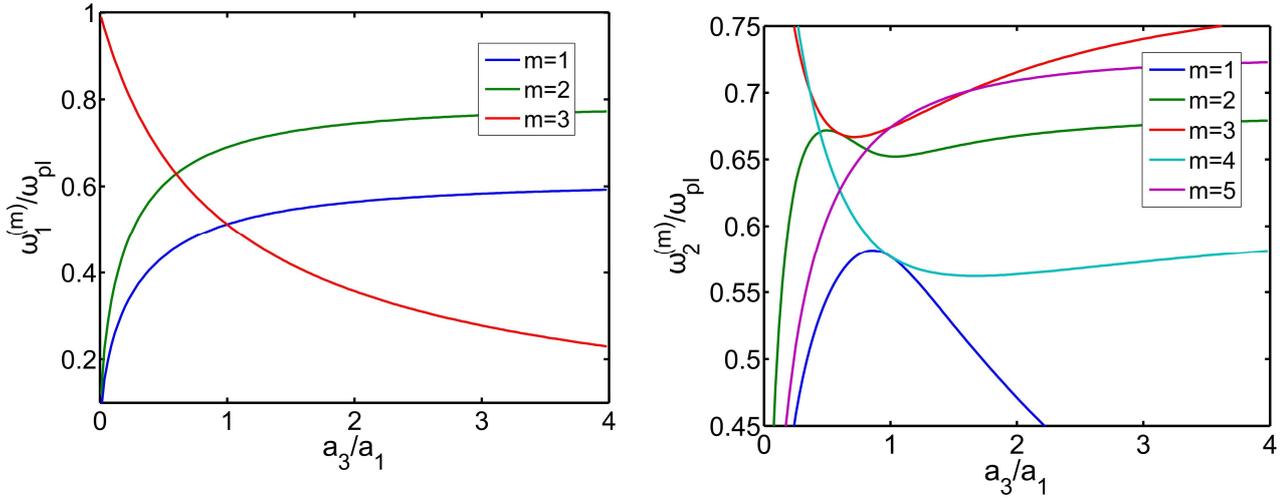

Fig.3. Eigenfrequencies of plasmonic oscillations of a triaxial ellipsoid as a function of the ratio of the semiaxes $a_3/a_1$ (for $a_2/a_1 = 0.6$). Solutions are shown in the following cases: a) $n=1$; b) $n=2$ [90].

The equations (41)(42) exhaust the solution of the problem of plasmon oscillations in a triaxial nano-ellipsoid in the quasi-static (Rayleigh) approximation. Corrections due to retardation effects and explicit expressions for the eigenfunctions of ellipsoidal resonators in Cartesian coordinates can be found in [90].

The effect of the mode structure in a silver nanospheroid on the temporal dynamics of plasmon oscillations in it was considered in [93], where it was shown that the use of only two quasi-normal modes to describe such a nanoparticle explains well the experimental data on measuring ultrafast dynamics using photoemission electron microscopy (PEEM) [94].

3.3. Optical Properties of Clusters of Plasmonic Resonators

Above, we have presented the results of works on plasmonic resonances in isolated particles that are topologically equivalent to a sphere. Even more interesting resonant oscillations are possible in nanoparticle clusters, since they have a more complex geometry and a larger number of control parameters.

In [95-97], an analytical description of the resonant properties of oscillations in a dimer of two spherical nanoparticles was developed.

Figure 4 shows the dependences of eigenfrequencies of transverse oscillations ($m=1$) of a plasmonic dimer on the distance between nanoparticles. The *L*- and *T*-modes are associated with the hybridization of the modes of individual nanoparticles [98,99], while the high-frequency *M*-modes, which appear only at short distances $R_{12}/(2R_0) \leq 1.2$, have no analog and correspond to bound states that arise in the strong interaction regime.

For the eigenfrequencies of the *M*-modes, the analytical solution has the form [95-97]:

$$\omega = \frac{\omega_{pl}}{\sqrt{1-\varepsilon_m^{(M)}}}; \varepsilon_m^{(M)} = -(m+M+\delta_m)\operatorname{arsinh}\sqrt{\frac{R_{12}^2}{4R_0^2}-1} \qquad (44)$$

$$(m=0,1,2....; M=1,2,3,......)$$

The corrections $\delta_m$ are independent of the mode number *M* and are equal to [1/2, -0.08578, -0.2639, -0.33, -0.3769, -0.4, -0.4172, 0.4289, -0.4377, -0.4446] for $m = 0…9$, respectively. These expressions show that, as the nanospheres approach each other, the resonant frequencies of the *M*-modes of the dimer tend to the plasma frequency $\omega_{pl}$.

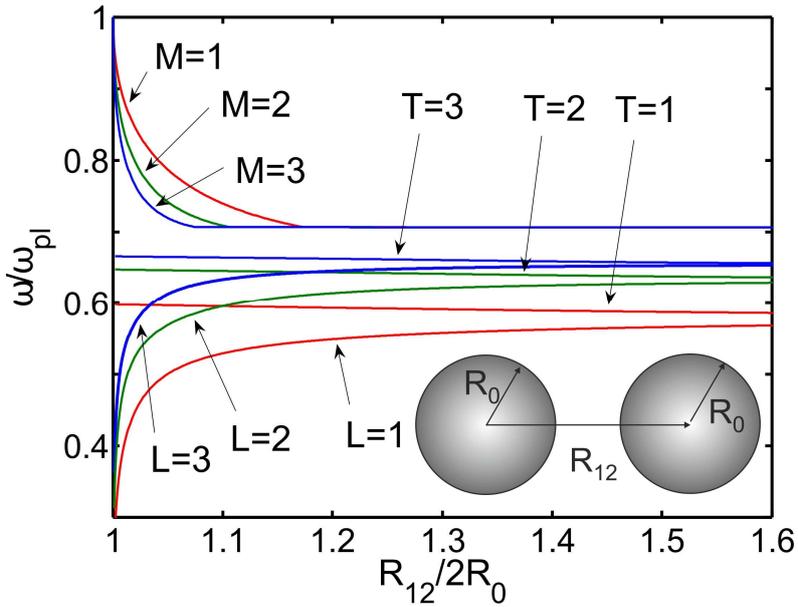

Fig.4. Eigenfrequencies of transverse plasmon oscillations ($m=1$) in a cluster of two nanospheres depending on the distance between them. The upper part of the figure shows clearly the emergence of new modes (*M*-modes) at small distances between the spheres. It is assumed that the permittivity is described by the Drude law (43).

For *L*-modes, there is a similar analytical solution [95-97]:

$$\omega = \frac{\omega_{pl}}{\sqrt{1-\varepsilon_m^{(L)}}}; \varepsilon_m^{(L)} = -\left(m+L-1/2\right)^{-1}/\operatorname{arsinh}\sqrt{\frac{R_{12}^2}{4R_0^2}-1}+... \quad (45)$$

$$(m=0,1,2,...; \quad L=1,2,...)$$

that is, as the nanospheres approach each other, the resonant frequencies of the $L$-modes of the dimer tend to zero.

Figure 5 shows the spatial distribution of the potential of the longitudinal $M$-, $L$-, $T$-modes ($m=0$). It can be seen from Fig.5 that the spatial structure of antisymmetric ($L$-modes) and symmetric ($T$-modes) wave functions generally corresponds to the structure of wave functions of isolated plasmonic nanoresonators. Namely, a positive charge is located on one hemisphere, while a negative charge equal to it due to the electroneutrality of the sphere is located on the opposite side of the sphere. In this case, the interaction between plasmonic nanoresonators is reduced to some quantitative redistribution of the charge on opposite hemispheres only.

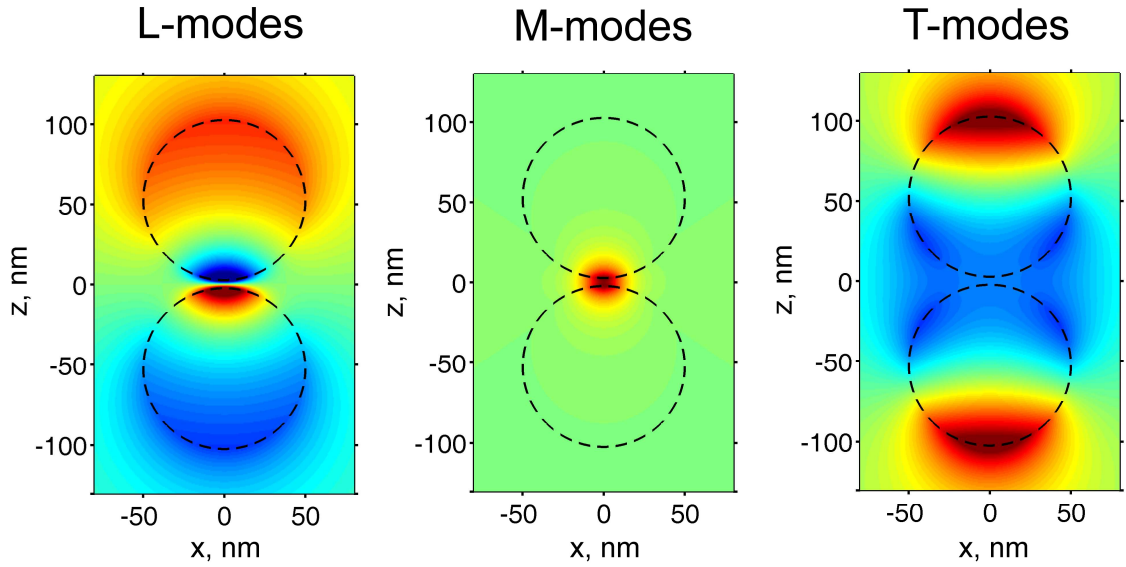

Fig.5. Spatial distribution of the potential in some plasmon modes in a cluster of two nanospheres ($m=0$, axisymmetric longitudinal oscillations) [97].

In the case of symmetric $M$-modes, the situation is different, and both positive and negative charges are concentrated near the gap between the nanospheres. Actually, at points far from the gap of the nanospheres, the wave functions of $M$-modes vanish.

As the distance between the spheres increases, the localization of the $M$-modes decreases and, at a critical distance between the spheres, these modes disappear, while the hybridized $L$- and $T$-modes do not experience significant changes.

The difference in the localization of the *M*-modes and *L*-, *T*- modes determines their fundamental difference with respect to the excitation fields. *L*-,*T*-modes have a polarizability of the order of the volume of a nanosphere $\alpha \sim R_0^3$ and effectively interact with uniform external fields of the appropriate orientation and symmetry. On the contrary, *M*-modes have a relatively low polarizability $\alpha \sim \Delta^3$, where $\Delta$ is the gap width between the spheres. Because of this, the *M*-modes are weakly (compared to the *L*- and *T*- modes) excited by uniform optical fields. On the other hand, *M*-modes interact effectively with strongly inhomogeneous fields that are localized near the gap between the spheres. Fields of this kind arise during the radiation of atoms and molecules located near the gap. This circumstance makes the *M*-modes extremely promising from the point of view of creating nanosensors and elements of nanodevices that are sensitive to the radiation of individual molecules.

A similar study of the resonance properties of a cluster of two plasmonic nanospheroids was carried out in [100].

Currently, plasmonic resonant nanostructures based on DNA origami technology are being actively studied [101-109]. This technology is very promising, as it allows one to arrange nanoparticles in a nanostructure with an accuracy higher than the accuracy of modern optical or electronic nanolithography. The creation of complex three-dimensional nanostructures from nanoparticles using DNA tethers is possible based on the canonical pairing of DNA bases according to Watson-Crick [101]. This approach opens unprecedented possibilities for the control of three-dimensional macromolecular structures on large scales. Using this technology, nanostructures of nanoparticles in the form of arrays [102-103], spirals [104], tetrahedra [105], chains [106-107], or nanorings [108] have already been realized. Recently, the possibility of transmitting information with low losses through a plasmonic waveguide made using DNA origami technology has been demonstrated [109].

In [110], the dependence of the resonant optical properties of Platonic clusters that can be synthesized using DNA origami technology on their topology and size was studied (Fig. 6).

To describe the optical properties of such clusters, a model was used in [110] where individual nanoparticles are characterized by isotropic polarizabilities and induced dipole moments with the interaction described by the retarded Green's function. Figure 7 shows the extinction spectra of various Platonic clusters with the same edge length $L_0$ and the same radius $R_0$ of silver particles, but with different topologies.

Comparison of Fig. 7a and Fig. 7b shows that the number of observed modes of Platonic clusters effectively increases as radiation and Joule losses decrease. Therefore, Fig. 7 can serve as a guideline for understanding the optical properties of meta-atoms in the form of Platonic clusters and interpretation of experimental data.

In [111], a plasmonic nanoresonator was synthesized in the form of a dodecapod consisting of a quartz core and 20 gold satellites - "pods" (Fig. 8), and it was experimentally shown that the value of the magnetic dipole moment induced in it is about 14% of the induced electric dipole moment, that is, optical magnetism has been demonstrated.

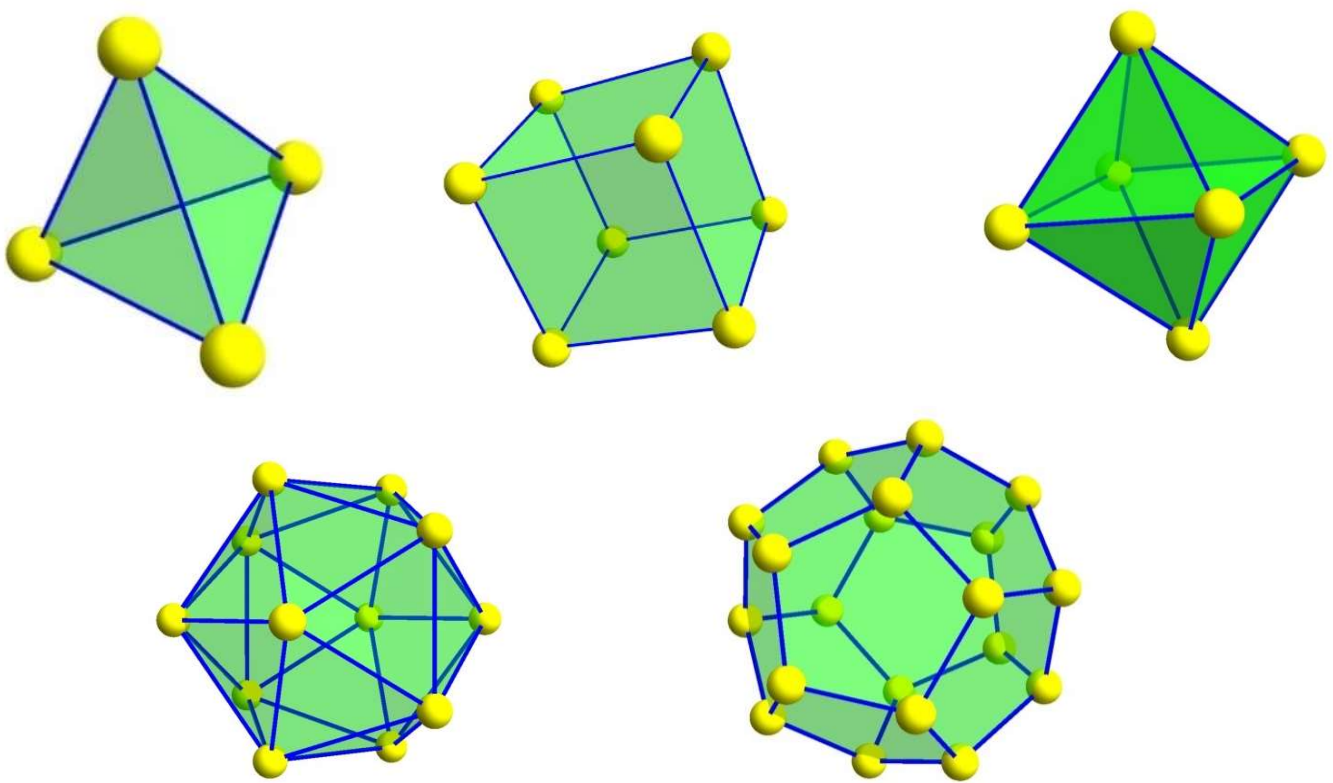

Fig.6. Platonic clusters of nanoparticles, which are shown as yellow balls.

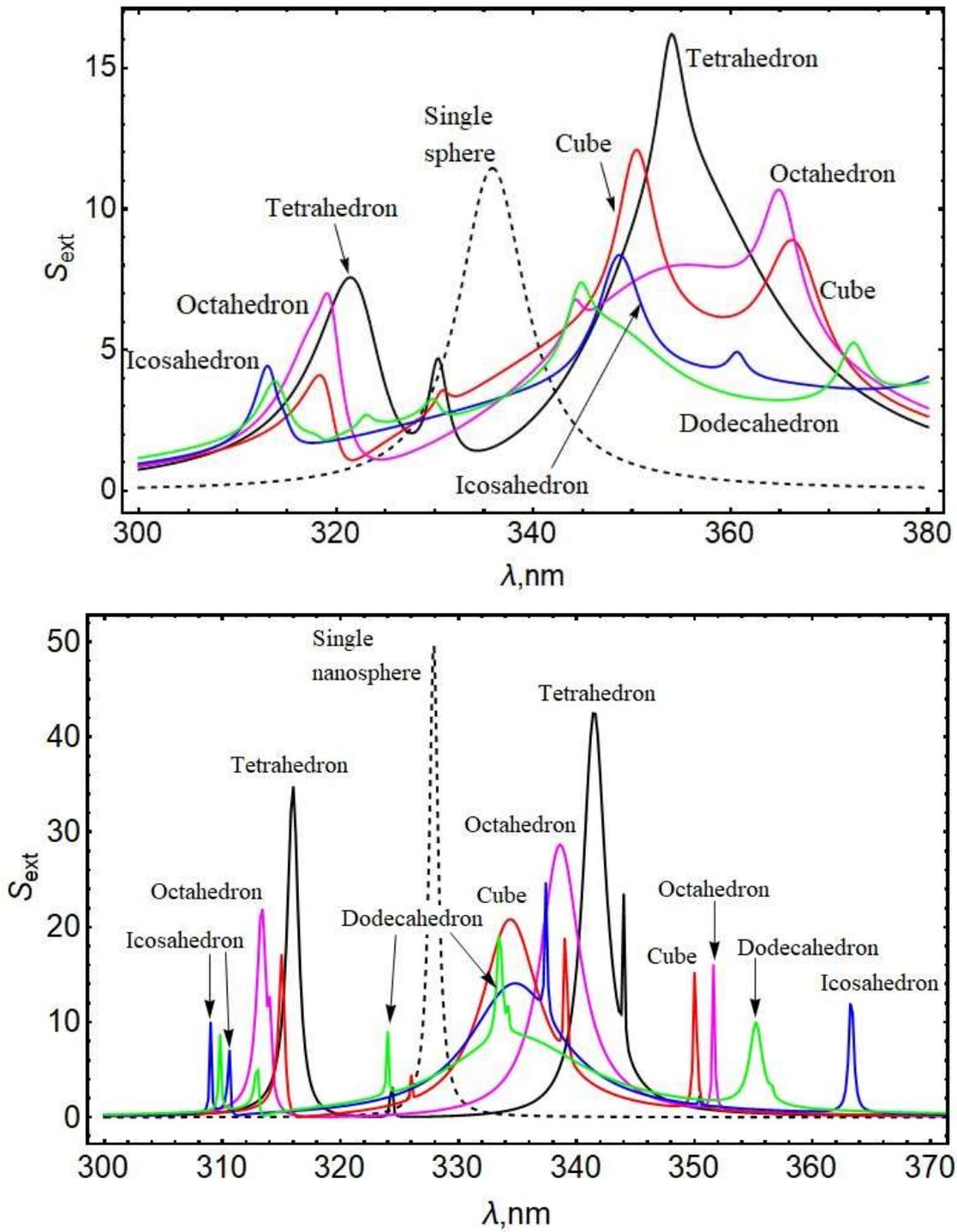

Fig. 7. Comparison of extinction spectra of Platonic clusters a) Ag, $L_0 = 50$ and $R_0 = 20$ nm and b) $L_0 = 25$ nm and $R_0 = 10$ nm (losses in the metal are reduced by a factor of 20 compared to real Ag) [110].

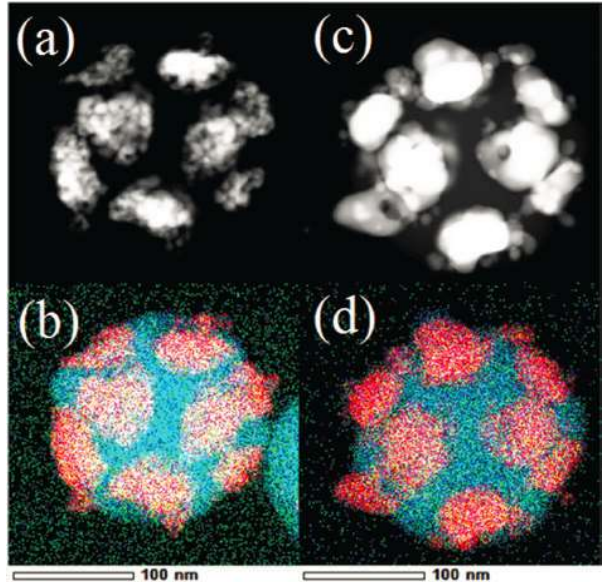

Fig. 8: Synthesized dodecapods. SEM image of the dodecapods before (a) and after (c) condensation of seed nuclei. b) and d) show the elemental composition of the synthesized dodecapod: blue color - $SiO_2$, red color - Au [111].

## 4. Modes in Dielectric Resonators

The main advantage of the plasmonic nanoresonators considered in the previous section is that natural oscillations can exist at arbitrary small sizes of nanoresonators. However, the value of this fact is significantly reduced by the circumstance that the Joule losses in plasmonic resonators are quite large and, in fact, the quality factors of plasmonic nanoresonators are limited to values of the order of 10 - 100. Recently, a research has begun on an alternative approach to controlling light at the nanoscale. It is based on dielectric nanoresonators, that is, nanostructures made of dielectric materials with a high or moderate refractive index, such as Si, Ge, GaAs, $TiO_2$, $Cu_2O$, etc. and very small internal (Joule) losses. Although most dielectric nanoresonators are larger than 100 nm in one or several dimensions, they are referred to in most publications as nanospheres, nanorods, nanodisks, and so on according to their shape. We will also call them nanoresonators, since their dimensions are much smaller than the wavelength in free space.

The history of dielectric resonators began with the work of 1939 [47], which laid the foundations for the theory of such resonators. An important development in this direction was given by van Bladel [74], who developed the theory of resonators with a large permittivity and noted the importance of the axial symmetry of such resonators for obtaining a high quality oscillations. Dielectric resonators were developed further in microwave technology [112,113,75].At present, in connection with the development of nanotechnologies, the elaboration of dielectric resonators in the optical range has also become possible [16,114,115 ].

4.1 Spherical Dielectric Nanoresonators

*4.1.1 usual (quasi-normal) modes*

As in the case of spherical plasmonic resonators, modes of spherical dielectric resonators are completely described by Mie theory, that is, the scattering coefficients (33), (34). On Fig. 9 the dependence of the absolute value of the Mie coefficients $q_4$ (33) on the permittivity and size of the sphere is shown.

It can be seen from this figure that, for subwavelength spheres, the resonance is possible only at high permittivities, when the wavelength in the dielectric becomes much smaller than the nanoresonator dimensions.

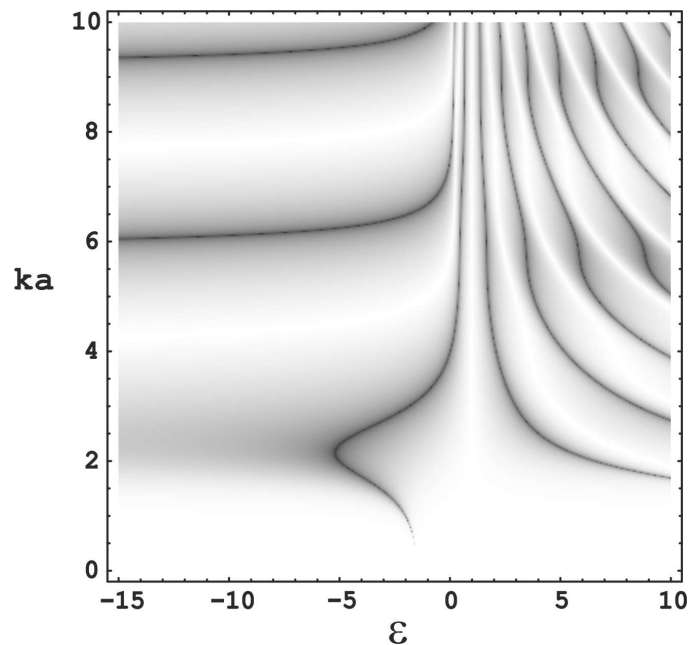

Fig.9. Dependence of the absolute value of the Mie coefficients $|q_4|$ (33) on the permittivity and the size parameter $k_0 a$ of a non-magnetic sphere located in vacuum. Black color corresponds to Mie factor equal to 1, i.e. resonance, while white color corresponds to Mie factor equal to 0 (no scattering) [1].

The poles of the Mie scattering coefficients at positive permittivity completely determine the eigenfrequencies of the quasi-normal modes of dielectric spheres. In the case of large permittivity $\varepsilon \rightarrow \infty$, the resonant properties of the dielectric sphere were studied in detail in [74,116, 117]. In the case of TM modes, the approach [117] leads to the following asymptotic expressions for the eigenfrequencies and Q-factors of the three lowest modes:

$$k_n^{TM} a = \frac{X_n^{TM}}{\sqrt{\varepsilon}}\left(1 - \frac{1}{n\varepsilon} + ..\right); Q_1^{TM} = \frac{\varepsilon^{5/2}}{2\left(X_1^{TM}\right)^3}; Q_2^{TM} = \frac{18\varepsilon^{7/2}}{\left(X_2^{TM}\right)^5}; Q_3^{TM} = \frac{2025\varepsilon^{9/2}}{2\left(X_3^{TM}\right)^7} \quad (46)$$

$$J_{n+1/2}\left(X_n^{TM}\right) = 0$$

For TE modes, the approach [117] leads to the expressions:

$$k_n^{TE} a = \frac{X_n^{TE}}{\sqrt{\varepsilon}}\left(1 - \frac{1}{(2n+1)\varepsilon} + ...\right); Q_n^{TE} = \frac{\Gamma(n+1/2)^2 (4\varepsilon)^{n+1/2}}{4\pi\left(X_n^{TE}\right)^{2n-1}}; J_{n-1/2}\left(X_n^{TE}\right) = 0 \quad (47)$$

Comparing the Q-factors of TM and TE modes, one can see that the Q-factors of TM modes are much higher. This is due to the fact that, as shown in [74], in resonators with a large permittivity, TM modes are confined, that is, in the limit $\varepsilon \to \infty$, the magnetic field is strictly confined inside the resonator, and the electric field is zero everywhere. TE modes, on the contrary, are not confined, and even in the limit $\varepsilon \to \infty$ the magnetic field of TE modes is non-zero outside the sphere. This circumstance leads to the fact that non-confined TE modes radiate more effectively, and therefore have lower quality factors. This circumstance is general.

*4.1.2. perfect nonradiating modes*

For a spherical non-magnetic nanoresonator of the radius $R$ in vacuum, the expression for the magnetic field of the perfect nonradiating TM modes has the form [40]:

$$H_\varphi^{(n)} = j_n(z_0) j_n\left(k_0\sqrt{\varepsilon}r\right) P_n^1(\cos\theta), r < R$$
$$H_\varphi^{(n)} = j_n(z_1) j_n(k_0 r) P_n^1(\cos\theta), r > R \quad (48)$$
$$z_0 = k_0 R; z_1 = k_0\sqrt{\varepsilon}R$$

which follows from the direct solution of Maxwell's equation.

Note that in the case of a sphere, the condition for the existence of perfect non-radiating modes (48)

$$\varepsilon j_n(z_1)\left[z_0 j_n(z_0)\right]' = j_n^{(1)}(z_0)\left[z_1 j_n(z_1)\right]' \quad (49)$$

coincides with the vanishing of the numerator of the Mie scattering coefficient. The dispersion equation (49), along with complex roots, also has real roots, which correspond to perfect non-radiating modes. Figure 10 shows the dependences of $\text{Re}(H_\varphi(r,\theta=\pi/2))$ on the radius for the perfect nonradiative $PTM_{101}$ mode and the usual quasi-normal $TM_{101}$ mode with radiative loss in a sphere with $\varepsilon=10$

From Fig.10, it can be seen that the field of the usual (quasi-normal) mode increases exponentially at infinity, while the field of the perfect nonradiating mode decreases and has no radiation losses! It was also shown in [40] that Joule losses in a dielectric have practically no effect on the properties of perfect nonradiating modes.

*4.1.3. Anapole current distributions*

In [32-36], the case of vanishing individual Mie scattering coefficients was studied, but in these works the main attention was paid to demonstrating how the expansion over Cartesian multipoles with taking into account toroidal moments is consistent with the expansion over vector spherical harmonics, thus leading to a zero field outside the spheres and, accordingly, to zero radiation. In these works, the distribution of fields inside the sphere, which leads to the compensation of the Cartesian electric dipole moment by the Cartesian toroidal electric moment, is called the anapole mode quite incorrectly, since the modes, by definition, are solutions of the Maxwell equation in the absence of currents. From our point of view, it is more correct to talk about anapole current distributions.

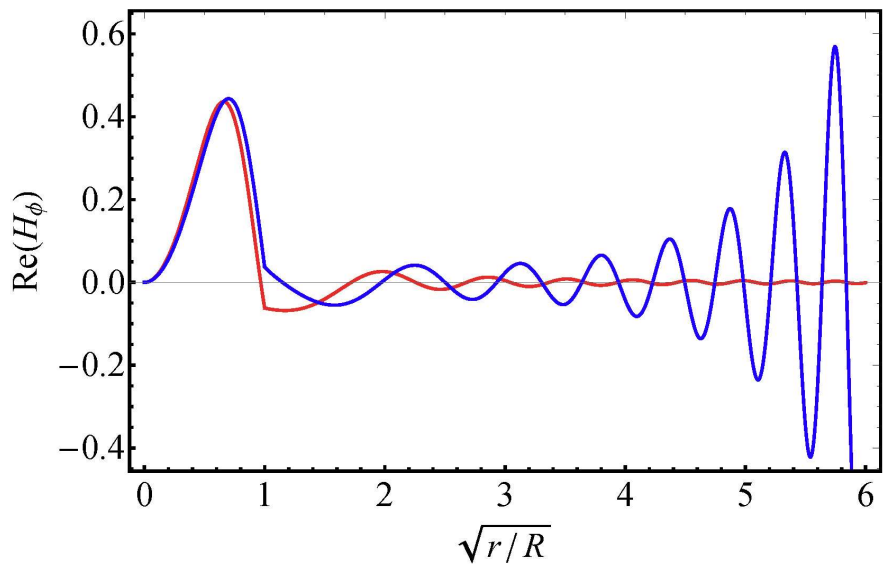

Fig.10. The dependence of $\text{Re}(H_\varphi(r,\theta=\pi/2))$ on the radius for an usual quasi-normal $TM_{101}$ mode (blue curve, $k_0R$=1.35715 - 0.160978$i$ ) and for the perfect nonradiating $PTM_{101}$ mode (red curve, $k_0R$=1.51893) in a sphere of radius $R$ with $\varepsilon$=10 [40].

*4.1.4. Excitation of Modes of Spherical Nanoresonators*

To understand the practical importance of usual quasi-normal and perfect nonradiating modes, one must also consider how they can manifest themselves in practical conditions, that is, find the conditions for their excitation. There are two lines of research here. The first deals with the conditions for excitation of several resonant modes simultaneously and the resulting interference phenomena in the far zone. For example, in [118] it was first noticed that at $2\pi\sqrt{\varepsilon}R/\lambda_0 > 2.7$, the radiation pattern of light scattered by a dielectric sphere of the radius $R$ transforms from Rayleigh non-directional scattering to forward scattering. In this paper, this effect was explained by the interference of electric and magnetic dipole moments induced in the sphere.

In [9-11], the results of [118] were generalized to the case of excitation of silicon spherical nanoresonators by radiation from a dipole source. It was shown that both electric and magnetic dipole moments can be simultaneously excited in Si nanoresonator, and the interference of their fields can lead to directional radiation due to the complete suppression of backward scattering (Huygens element) even from a single nanoparticle [9,11,119]. In a Si nanoparticle with the radius of 65 nm, this effect is achieved, since in a certain frequency range ($\lambda$=570 nm) it is possible to ensure that the electric and magnetic dipole moments induced in the particle are the same in amplitude and phase. In this case, the radiation of the dipole + nanosphere system will be directed from the dipole to the side where the nanoparticle is located. At $\lambda$=490 nm, the electric and magnetic dipole moments induced in the same particle will have the same amplitudes, but the phases will be shifted by 1.3 radians, leading to a change in the direction of radiation of the system under consideration to the opposite. The importance of the contribution of induced magnetic dipoles to light scattering led even to the appearance of the term "magnetic light" [120].

Another line of the research of the manifestation of optical resonances in nanoparticles is, on the contrary, the detection of only one resonant mode, quasi-normal or perfect nonradiating one. This is a more difficult task, since specially prepared light beams are needed for this, since always the sphere excitation by a standard plane wave conceals subtle effects associated with individual modes by strong fields from excitation of a continuum of quasi-normal modes.

In [39], an exact analytical solution was found for the problem of excitation of a dielectric sphere by an axisymmetric Bessel beam

$$H_\varphi = H_0 J_1(k_0 \rho \sin\beta) e^{ik_0 z \cos\beta} \text{ (TM polarization)}$$
$$E_\varphi = E_0 J_1(k_0 \rho \sin\beta) e^{ik_0 z \cos\beta} \text{ (TE polarization)}$$
(50)

where $\rho$ and $z$ are cylindrical coordinates, and $\beta$ is the conical angle of the beam. Such a beam has a complex spatial structure (see Fig. 13), enabling one to control its interaction with axisymmetric particles effectively.

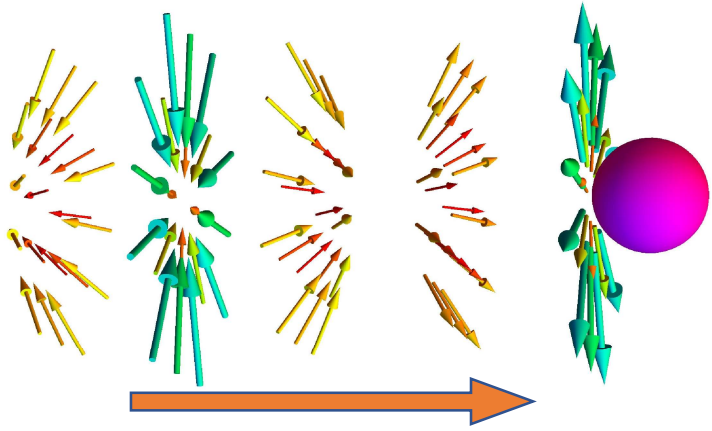

Fig.11. Structure of the electric field in an axisymmetric TM Bessel beam.

The expression for the magnetic field scattered by the sphere in spherical coordinates $(r,\theta,\varphi)$ has the form [39]:

$$H_\varphi^R(r,\theta) = iH_0 \sum_{n=1}^{\infty} i^n q_n \frac{2n+1}{n(n+1)} h_n^{(1)}(k_0 r) P_n^1(\theta) P_n^1(\beta)$$
(51)

where $q_n$ are the Mie coefficients (33).

Solution (51) describes the excitation of a sphere (exactly) and other axisymmetric nanoparticles (qualitatively) by axisymmetric beams having only TM or TE polarization. This solution removes the limitation imposed by the Mie1908 solution, where the TM and TE modes cannot be separated. An important feature of the solution (51) is also that it nontrivially depends on the cone angle $\beta$, whose tuning can control the excitation or suppression of certain modes.

On Fig. 12 one can see the dependence of the generalized $Q$ factor (3), scattered power and stored energy on the size parameter $k_0 R$ of the nanosphere for a beam with TM polarization.

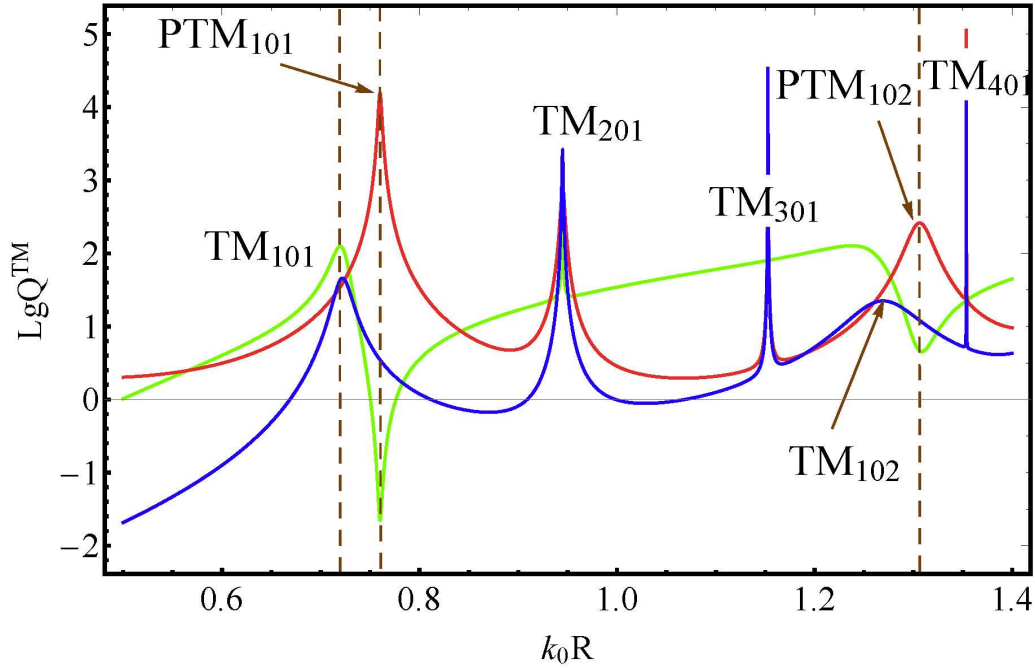

Fig. 12. Dependence of the generalized quality factor $Q$ (red curve, (3)), scattered power (green curve), and stored energy (blue curve) on the size parameter of the nanosphere $k_0R$ (axisymmetric Bessel beam, TM polarization, $\varepsilon = 36$, $\beta = \pi/4$) [39]).

Figure 12 clearly shows ultra-high-quality ($Q > 10^4$) perfect modes (PTM$_{101}$) with very low scattering, with the magnitude limited only by the presence of a continuum of usual quasi-normal modes. The positions of the perfect modes correspond exactly to the solutions of the dispersion equation (49).

The dependence of the found solution on the parameters of the Bessel beam makes it possible to find the conditions for the excitation of any preassigned usual or perfect mode with an unlimited radiation $Q$-factor.

Figure 13 shows the corresponding distribution of the magnetic field when the sphere is illuminated with a superposition of 4 Bessel beams with specially selected parameters, suppressing the lowest order modes radiation completely. As a result, only the extremely weak 32-field (triacontadipole) radiation survives. Note that inside the sphere the field distribution corresponds to the pure PTM$_{102}$ mode.

Recently, in [15], an attempt was made to study experimentally TM and TE resonances in Si nanospheres separately using radially and azimuthally polarized beams. Fig. 14 shows the results of such an experiment and the corresponding computer simulation.

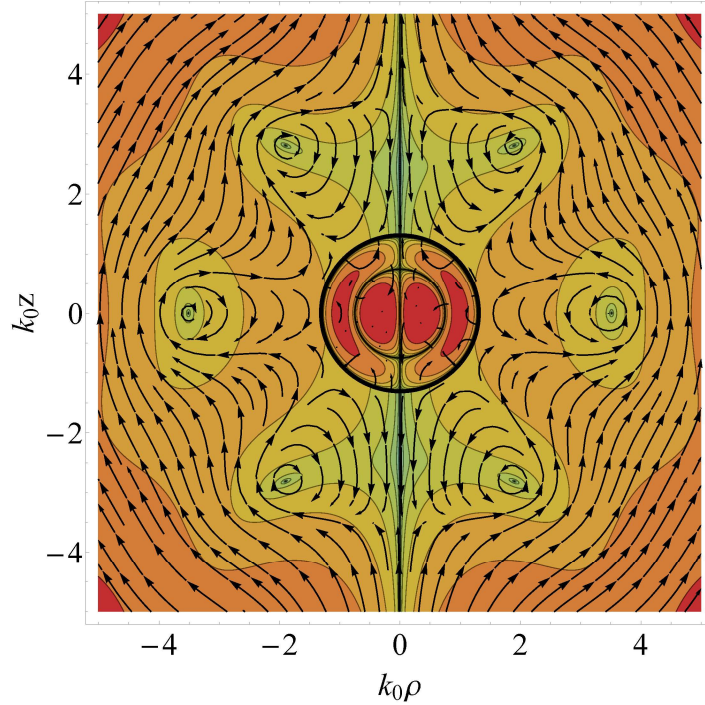

Fig. 13. Practically nonradiating 32-field distribution of the magnetic field when the sphere is excited by a superposition of 4 Bessel beams ($k_0R=1.31$, $\varepsilon = 36$) [39].

From Fig.14 it can be seen that for a radially polarized beam, mainly electric modes (both ED and EQ) are excited, while during excitation by an azimuthally polarized beam, mainly magnetic modes (both MD and MQ) are excited. Note that for the radial excitation there is a scattering minimum near $\lambda$=450 nm, which probably corresponds to the perfect nonradiative mode.

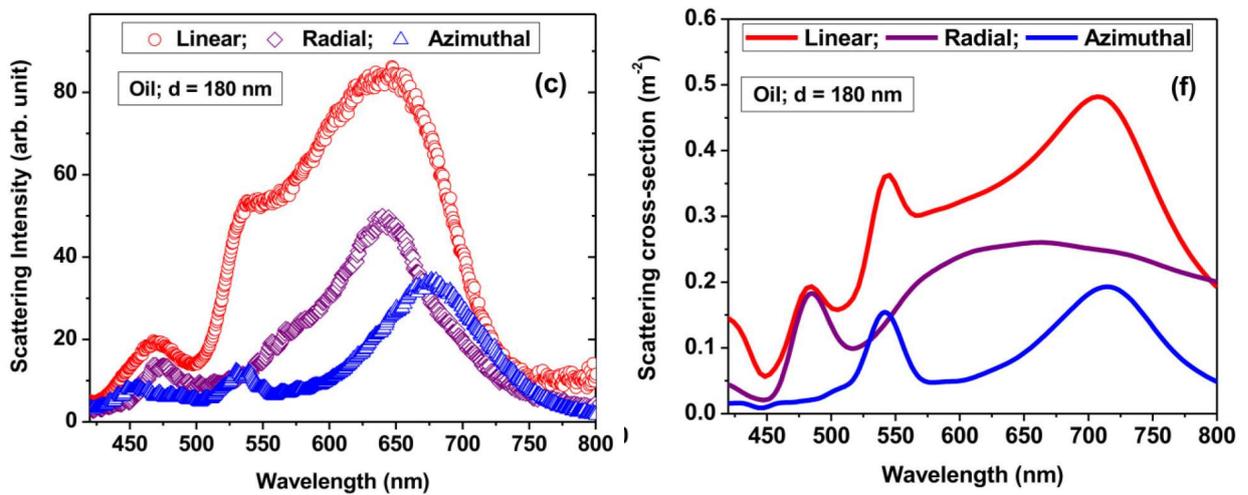

Fig.14. Selective excitation of multipole resonances using radially and azimuthally polarized beams. Experimental (a) and theoretical (b) spectra of single Si nanospheres with the diameter of 180 nm in immersion oil with $n$=1.48 [15].

### 4.2. Spheroidal Nanoresonators

#### 4.2.1. *Quasi-normal modes*

In principle, usual quasi-normal modes of spheroidal nanoresonators can be found on the basis of an analytical solution of the problem of plane wave scattering by a dielectric spheroid [121,122]. However, we failed to find such a solution in the literature.

The case of small deviations from the spherical shape was considered in [123], where the small deviations from the spherical shape were described by the expression:

$$r/R = 1 + \mu F(\theta, \varphi) \tag{52}$$

with the small parameter $\mu$, and the change in the quality factor of the mode associated with this perturbation was sought in the form of a series:

$$1/Q = 1/Q_0 + \mu C_1 + \mu^2 C_2 + ... \tag{53}$$

where $Q_0 \gg 1$ is the quality factor of the unperturbed sphere mode.

The main and most important result of [123] is that it shows that for any high-quality mode of the sphere ($Q_0 \to \infty$), $C_1 = O(1/Q_0) \to 0$, $C_2 \geq 0$, i.e. any small shape perturbation of the high-quality spherical resonator leads to a decrease in the quality factor of its modes. This result is quite consistent with the intuitive idea of a sphere as an perfect object that minimizes the scattering of waves on its surface. In the case of not very high $Q$-factors, the coefficient $C_1$ becomes non-zero, and the maximum $Q$-factor can be realized for particles with a shape slightly different from spherical.

In [61], the eigenfrequencies and $Q$-factors of the spheroidal resonator modes were found numerically by solving the Muller boundary integral equations discretized with the Nystrom method. In [61], a determinant was found to search for eigenmodes frequencies and, with its help, the resonant frequencies and quality factors of usual eigenmodes of a dielectric spheroid with semiaxes $a$ (along the axis of symmetry) and $b$ (perpendicular to the axis) were found. It was assumed that the spheroid had a volume equal to the volume of a sphere of the radius $R$ and its surface is described by the equation:

$$(\rho/b)^2 + (z/a)^2 = 1; a = Rt^{2/3}; b = Rt^{-1/3} \tag{54},$$

where $t = a/b$. For $t < 1$ we have an oblate spheroid, for $t > 1$ it is prolate.

Figure 15 shows the dependencies of the quality factors of the lowest modes for spheroids with $\varepsilon=38$ on the aspect ratio of the semiaxes.

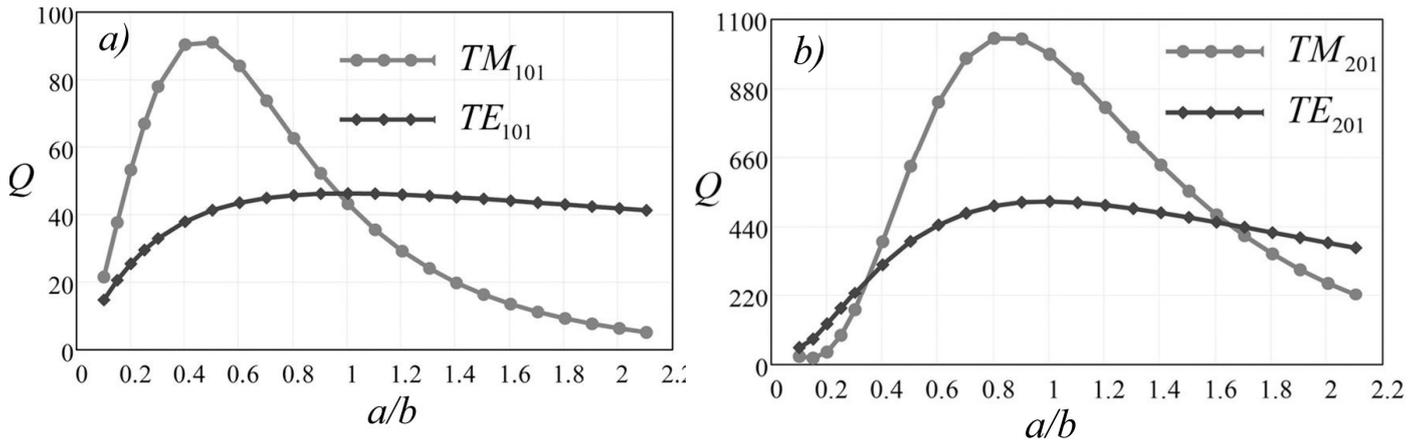

Fig. 15. Dependences of the quality factors of the lowest modes for spheroids with $\varepsilon=38$ on the aspect ratio of the semiaxes [61].

From Fig. 15, one can draw extremely important and highly non-trivial conclusions that:
1) TM modes have significantly higher quality factors than TE modes, due to the fact that they are confined modes (see (26) и (28));
2) The quality factors of the lower TM modes ($TM_{101}$, $TM_{201}$ modes) are maximal not for spheres, but for oblate spheroids. Higher-order modes have maximal $Q$-factors for spheres in accordance with the results of [123];
3) The quality factors of all TE modes are maximal for spheres, but not for oblate or prolate spheroids, which also agrees with the results of [123];
4) $TM_{101}$ (electric dipole) and $TE_{101}$ (magnetic dipole) mode resonances have good overlap for oblate spheroids ($a/b<1$).

The latter circumstance was used in [124] to maximize forward scattering by an oblate dielectric spheroid.

A more detailed study of the relationship between the real and imaginary parts of resonant frequencies was carried out in [125] using simulations with COMSOL Multiphysics [78]. Fig. 16 shows how the real and imaginary parts of the resonant frequency of a silicon spheroid change depending on the aspect ratio $a/b$ of the spheroids.

It can be seen from Fig. 16 that the trajectories of natural frequencies in the complex plane when the shape of the spheroid changes are quite complex, especially in the case of higher modes. Nevertheless, again the high-$Q$ modes without zeros in the radial direction ($TM_{301}$, $TM_{401}$, $TM_{501}$, $TM_{601}$, $TM_{701}$) have a near-zero imaginary part of the frequency and, therefore, the maximal *Q*-factor in the case of a spherical nanoparticle. The low-*Q* $TM_{101}$ mode has the maximal *Q* factor for oblate spheroids, agreeing with the results of [61] (see Fig. 16).

In [125], similar results were also obtained for TE modes in spheroids.

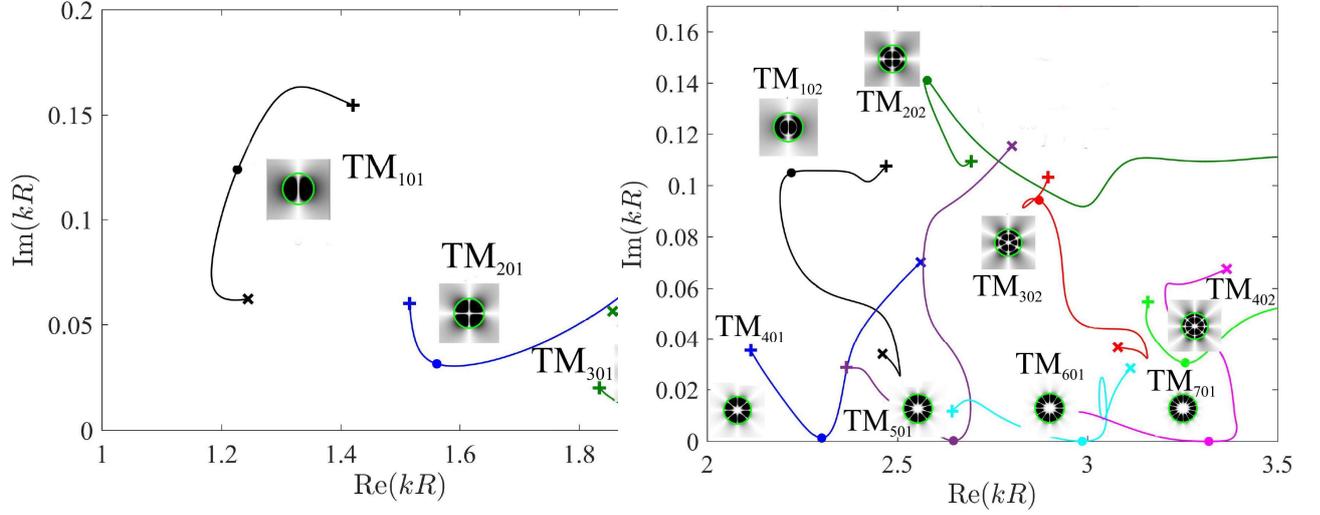

Fig.16. Dependence of the real and imaginary parts of the resonant frequencies of axisymmetric $TM_{n0m}$ modes on the shape of a silicon spheroid with $\varepsilon=12$. Markers "×", "+" and "•" on the curve correspond to the oblate spheroid (*a/b*=0.4), the prolate spheroid (*a/b*=1.6), and the sphere (*a/b*=1), respectively. The insets show the distributions of $H_\varphi$ fields in the modes, the numbers above the insets show the orbital and radial quantum numbers of the mode. *R* is the radius of a sphere with a volume equal to that of the spheroid. Adapted from [125].

*4.2.2. Perfect nonradiating modes*

Perfect nonradiative modes are not a feature of spherical geometry and definitely exist for axisymmetric bodies of an arbitrary shape. In [40], by expanding the solutions of Eqs. (14) (15) in terms of radial $S_{n1}(c,\xi)$ and angular $PS_{n1}(c,\eta)$ spheroidal wave functions [122], it was shown rigorously that such modes exist for arbitrary spheroids, and their eigenfunctions and frequencies were found.

The dispersion equation describing perfect nonradiating modes for prolate spheroids with permittivity $\varepsilon$ has the form [40]:

$$DetM = 0$$
$$M_{np} = \Pi_{np}(c_1,c_0)\left(\varepsilon SD_{p1}(c_0,\xi_0)S_{n1}(c_1,\xi_0) - S_{p1}(c_0,\xi_0)SD_{n1}(c_1,\xi_0)\right) \quad (55)$$

where

$$\Pi_{n,p}(c_1,c_0) = \int_{-1}^{1} d\eta \, PS_{n1}(c_1,\eta) PS_{p1}(c_0,\eta)$$

$$SD_{n1}(c,\xi_0) = \frac{\partial \left(\xi_0^2 - 1\right)^{1/2} S_{p1}(c,\xi_0)}{\partial \xi_0} \tag{56}$$

$$\xi_0 = t/\sqrt{t^2 - 1}; \quad c_0 = k_0 R \sqrt{t^2 - 1}/t^{1/3}, \quad c_1 = c_0 \sqrt{\varepsilon}$$

In (56), $k_0 = \omega/c$, and $R$ is the radius of the sphere, which defines the fixed volume of the spheroid. Dispersion equation (55) is also valid for an oblate spheroid after the corresponding analytic continuation.

We emphasize once again that the fields of perfect nonradiating modes found in [40], are not equal to zero outside the spheroidal resonator, that is, these are not "anapole current distributions" with the field equal to zero outside the resonator. It is also important that perfect nonradiating modes exist for any shape of a spheroid!

The found perfect nonradiating modes of spheroids are also not abstract solutions, they are of great importance for finding the conditions under which the scattered power becomes minimal or even zero. To demonstrate the practical importance of perfect nonradiating modes, the characteristics of scattering of an axisymmetric Bessel beam

$$H_\varphi \sim J_1(k_0 \sin\beta \rho) \cos(k_0 z \cos\beta) \tag{57}$$

by spheroidal resonators of various shapes were calculated in [40]

Figure 17 shows the dependence of the scattered power, the energy stored in the spheroid and the quality factor on the size parameter of nanospheroids with $a/b = 0.7$.

It can be seen from this figure that, under such excitation, all scattering minima and $Q$-factor maxima ($Q > 10^5$) are due to perfect nonradiating modes. The positions of these modes correspond exactly to the solutions of the dispersion equation (55). The finite quality factor of perfect modes (which are infinite in theory) is due to the fact that there is a continuum of usual quasi-normal modes that result in small but finite radiation, which can be reduced unlimitedly by optimal tuning of a beam. In any case, the power scattered by perfect nonradiating modes (PTM$_{101}$ and PTM$_{301}$) is 4 to 5 orders of magnitude less than the power scattered by usual quasi-normal modes (TM$_{101}$ and TM$_{301}$). Accordingly, the quality factors of perfect non-radiating modes are 4–5 orders of magnitude greater than the quality factors of usual modes.

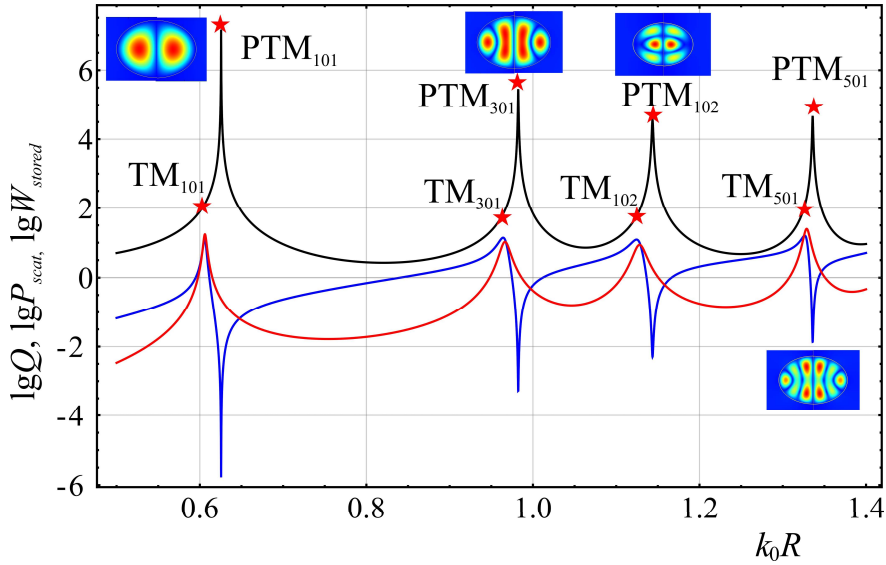

Fig.17. Dependence of scattered power (blue curve), stored energy (red curve), and quality factor (black curve,(3) ) on the size parameter of nanoparticles. TM symmetric excitation (57) with $\beta = \pi/4$, $\varepsilon = 50$, $a/b = 0.7$. All maxima of $Q$ factor correspond to perfect nonradiating modes. The asterisks indicate the $Q$ factors of usual modes ($TM_{101}$, $TM_{301}$, $TM_{501}$) and perfect nonradiating modes ($PTM_{101}$, $PTM_{301}$, $PTM_{501}$)[40].

The existence of TE perfect nonradiative modes in spheroids was also theoretically demonstrated in [40].

### 4.3. Dielectric Nanoresonators of More Complex Shapes
*4.3.1. Usual modes of finite-height cylinders*

Relatively simple shapes of resonators considered above can be described theoretically in details and, consequently, the physics of processes in such nanoresonators is quite understandable. However, from a technological point of view, such resonators are difficult to implement, especially in optical integrated circuits. Therefore, in recent years, special attention has been paid to the study of cylindrical dielectric nanoresonators, that is, the resonators with external size smaller than the wavelength in the air.

For circular cylindrical resonators of finite height, a lot of theoretical and experimental researches have been done, and therefore their characteristics are generally well known. In particular, Table 1 [75] lists the characteristics of some high-$Q$ modes of cylindrical nanocavities and the values of the index $P$ (see (28)) that determines their $Q$-factor as a function of the refractive index. In the same work, one can find corrections related to the shape of a particular cylinder.

**Table 1** [75]. Characteristics of some high-quality modes of cylindrical nanocavities and the values of the index $P$ (see (28)) that determines the dependence of their quality factors on the refractive index.

| mode | field structure inside the resonator | far field structure | multipole orientation | index $P$ |
|---|---|---|---|---|
| $TE_{01\delta}$ | $H_z = J_0(hr)\cos\beta z$ <br> $E_z = 0$ | magnetic dipole | $\parallel$ axis | 1.27 |
| $HE_{11\delta}$ | $E_z = J_1(hr)\cos\beta z \times e^{i\varphi}$ <br> $H_z = 0$ | magnetic dipole | $\perp$ axis | 1.30 |
| $HE_{21\delta}$ | $E_z = J_2(hr)\cos\beta z \times e^{i2\varphi}$ <br> $H_z = 0$ | magnetic quadrupole | $\perp$ axis | 2.49 |
| $EH_{11\delta}$ | $H_z = J_1(hr)\cos\beta z \times e^{i\varphi}$ <br> $E_z = 0$ | electric dipole | $\perp$ axis | 2.71 |
| $TE_{011+\delta}$ | $H_z = J_0(hr)\sin\beta z$ <br> $E_z = 0$ | magnetic quadrupole | $\parallel$ axis | 2.38 |

For example, for the mode $EH_{11\delta}$ ( electric dipole perpendicular to the axis) index $P = 2.71$, and this gives the value of the quality factor $Q=1.4\times10^5$ at $\varepsilon=80$!

A similar study was carried out in [126], where approximation formulas were also found for the eigenfrequencies and quality factors of several principal modes of dielectric cylindrical nanocavities. In particular, for $TM_{01\delta}$ (electric dipole along the axis), the approximation expression for the quality factor is:

$$Q_{TM_{01\delta}} = 0.008721\varepsilon^{0.888413}$$
$$e^{0.0397475\varepsilon_r}\left\{1-\left[0.3-0.2\left(\frac{a}{h}\right)\right]\left[\frac{38-\varepsilon}{28}\right]\right\} \times \qquad(58)$$
$$\left\{9.498196\left(\frac{a}{h}\right) + 2058.33\left(\frac{a}{h}\right)^{4.322261} e^{-3.50099(a/h)}\right\}$$

Figure 18 shows the $Q$-factors of the lowest $TE_{01\delta}$ and $TM_{01\delta}$ modes in a cylindrical resonator with $\varepsilon=38$ [61,127].

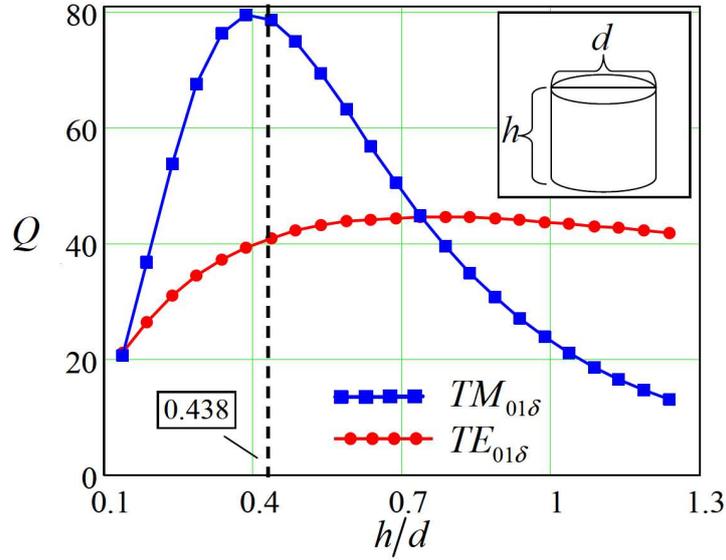

Fig.18. The dependence of $Q$-factors of the lowest $TE_{01\delta}$ and $TM_{01\delta}$ modes in a cylindrical resonator with $\varepsilon=38$ as a function of aspect ratio $h/d$ [61,127].

*4.3.2. "Supercavity" modes in cylinders*

Recently, much attention has been paid to the theoretical and experimental study of high-quality modes in cylindrical nanocavities [21,25,28,37,38,128]. In view of the high-quality factor, these, in general, usual quasi-normal modes have even been called "supercavity" modes or "quasi-BIC states". These modes arise at optimal cylinder shapes, at which the lowest spherical harmonic disappears in the multipole expansion of the eigenmode (18).

In [69], modes were studied in a more complex cylinder of finite height with a regular triangle at the base. In view of the presence of sharp corners, it is difficult to expect the presence of high-quality modes in such resonators in the general case. However, at certain parameters, there are modes (sometimes they also call them "supercavity" modes) that do not feel sharp corners, and their quality factors approach those of smoother resonators of the same volume, such as spheres or spheroids (see Section 4.2).

In [38], an experimental observation of "supercavity" modes in subwavelength ceramic resonators in the radio frequency range was carried out. Fig. 19 shows the scheme of the experiment and the results of measuring the scattering coefficient.

Mathematical analysis of the measured scattering coefficients showed that by the fine tuning of the shape of the resonator, in the experiment it is possible to achieve a substantial increase of the quality factor, reaching values up to $Q_{super} = 1.25 \times 10^4$ for a mode in ceramics with the loss tangent about $10^{-4}$. Direct finite element simulation shows that the intrinsic radiation $Q$-factor of the cylinder mode is about $Q_{super}$ =180000, approaching the $Q$-factor of the TE$_{401}$ mode in a sphere with the same permittivity and the same volume, $Q_{sph}$=213945, $k_0r$=1.04.

Figure 19c shows that the spatial structure of the "supercavity" mode is close to the structure of the field in a spherical resonator, and the cylinder edges are in the region of a small field, explaining the absence of scattering by them and the high-quality factor of these modes. "Supercavity" modes have a $Q$-factor lower than the $Q$-factors of modes of similar structure in spheres of the same material and the same volume, this is in full agreement with the results of [123] (see Section 4.2.1).

A system similar to that shown in Fig. 19a was also studied in [129] to see if it could achieve super scattering [130-132], rather than the maximal quality factor. By optimizing the relationship between the spherical multipole content of the excited mode, it was shown in [129] that the optimized structures in the limit of dipole scattering have a scattering cross section that exceeds the limit for a sphere by a factor of 4.

"Supercavity" modes in a system of two coaxial silicon cylinders (see Fig. 20) were analyzed in detail in [133] using simulations with COMSOL Multiphysics [78]. The physics of this problem is much richer than the physics of a single cylinder since an additional parameter $L$ - the distance between the centers of the cylinders - appears in it. Fig. 21 shows the evolution of the resonant frequencies of axisymmetric TE modes in the complex plane as a function of the distance between the cylinders.

Again, this figure shows that the minimum imaginary part of the frequency occurs at $L$=1.75 $h$ on the curve corresponding to antisymmetric hybridization. In this case, the quality factor is equal to $Q$ = 5500, and the field structure is close to the structure of the TE$_{602}$ mode in the sphere. Another high-quality mode appears at $L$=1.08$h$ on the curve corresponding to symmetrical hybridization. The structure of this mode is close to the structure of the TE$_{502}$ mode in a sphere, and the quality factor is lower than in the case of antisymmetric hybridization at $L$=1.75 $h$.

An interesting approach to the creation of high-$Q$ optical nanoresonators was proposed in [23], where it was shown that a cavity in a spherical or cylindrical resonator can significantly reduce the fraction of lower spherical harmonics, since they are localized closer to the cavity center and the cavity acts as a filter of lower harmonics. It is important that the contribution of electric multipoles decreases especially strongly. This interesting result was confirmed experimentally on silicon nanorings (see also Fig. 1).

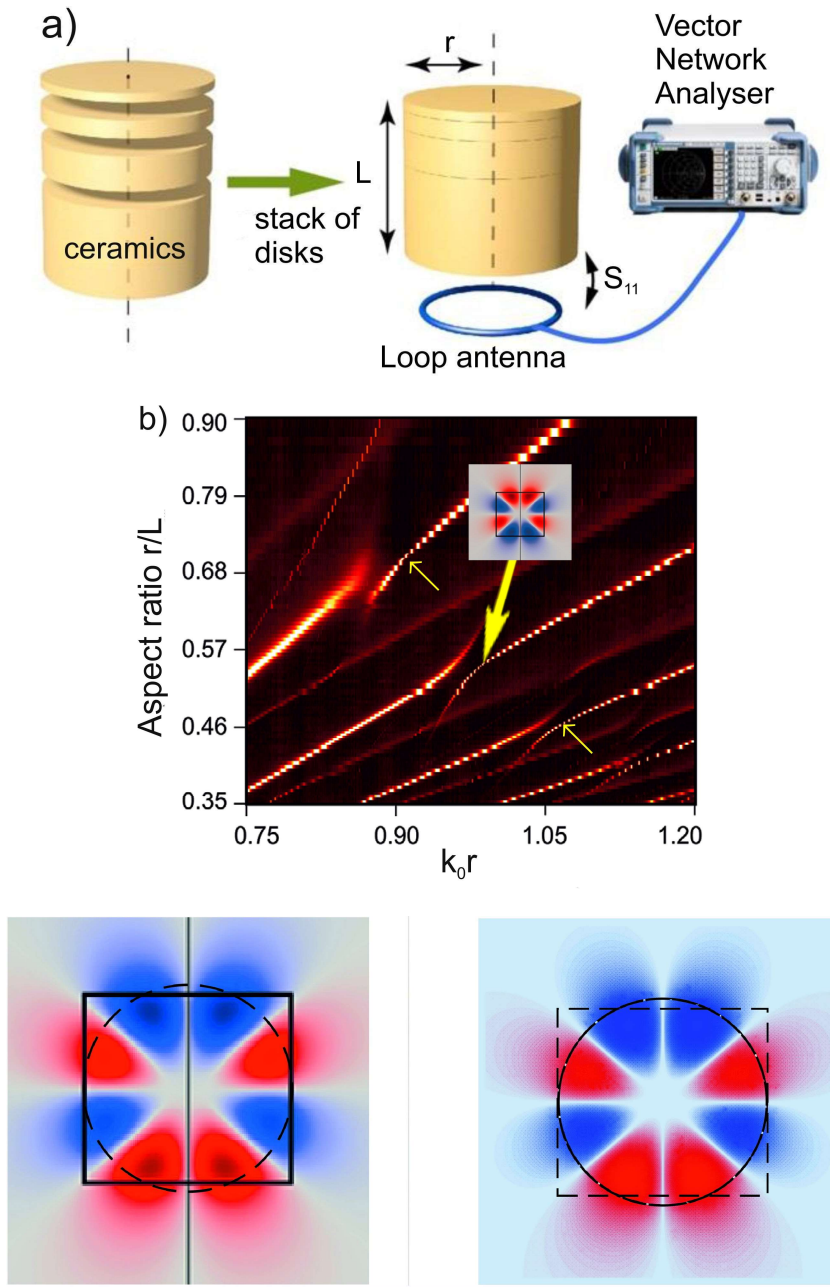

Fig. 19. Experimental observation of "supercavity" modes in subwavelength cylindrical resonators. (a) Experimental setup for measuring scattering in the radio frequency range. The resonator consists of several ceramic disks of different heights, combined in a single cylinder with the radius $r = 15.7$ mm, the dielectric constant $\varepsilon = 44.8$ and the ceramic loss tangent of about $\tan\delta = 10^{-4}$. The TE modes of the resonator are excited by a loop antenna connected to a vector network analyzer. (b) Dependence of the experimentally measured scattering coefficient $1 - |S_{11}|$ on the size parameter $k_0 r$ and the aspect ratio $r / L$. The arrows show the positions of the "supercavity" modes and the spatial structure of one of them (adapted from [38]); c) comparison of the structures of the "supercavity" mode in a cylinder ($\varepsilon = 44.8$, $Q=180000$) and the $TE_{401}$ mode in a sphere ($\varepsilon = 44.8$, $Q=214000$).

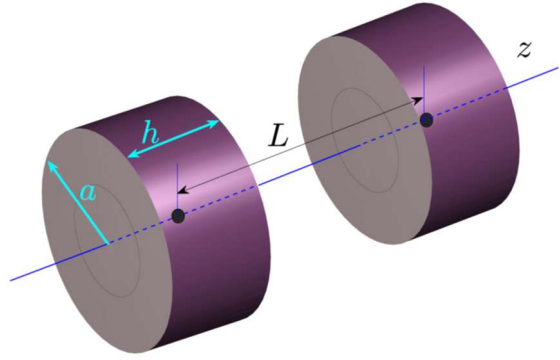

Fig.20. Geometry of the problem of "supercavity" modes in a system of two coaxial silicon cylinders.

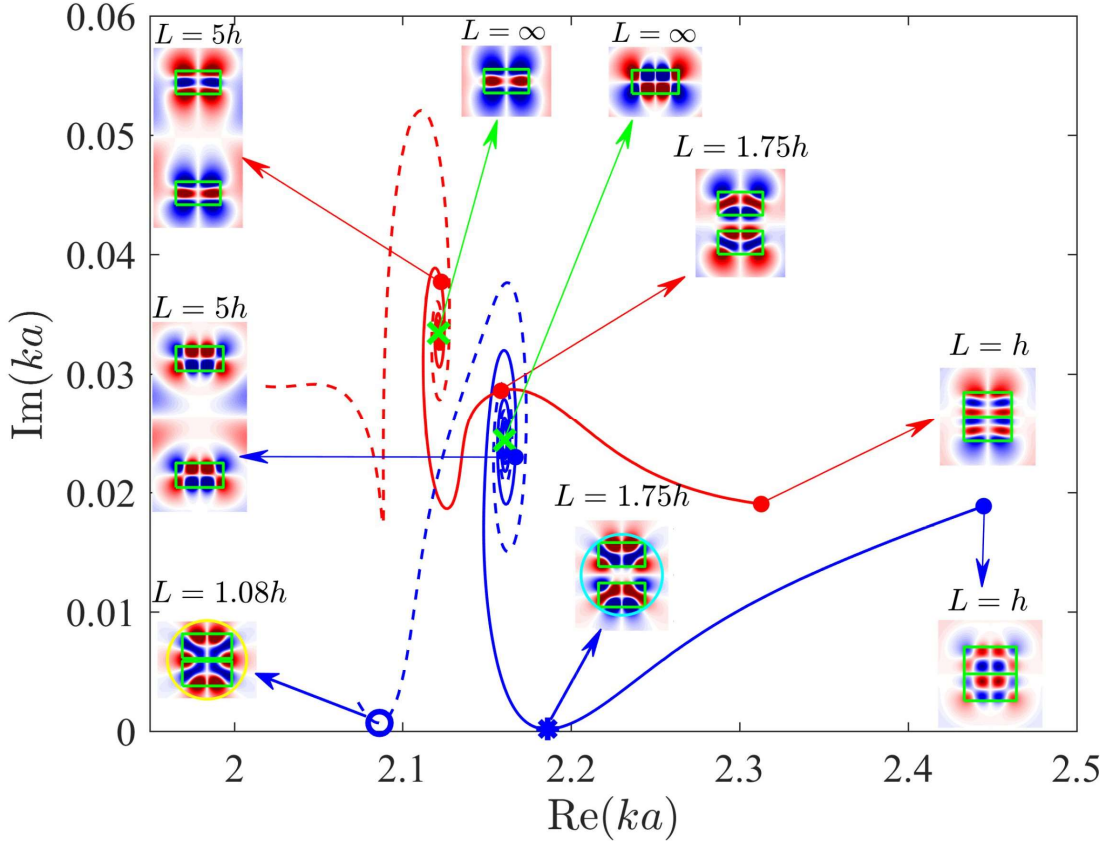

Fig.21. Evolution of the resonant frequencies of axisymmetric TE modes in the complex plane as a function of the distance between the centers of the cylinders $L$. The positions of the resonant frequencies of an isolated cylinder ($L = \infty$) are shown by green crosses ×. Trajectories of antisymmetric and symmetric modes beginning from the point $L = \infty$ are shown by solid and dashed lines. $a = 0.5$ μm, $a/h = 0.96$. The insets show the distribution $E_\varphi$ and the corresponding distance $L$ between the cylinders [133].

In general, it can be said that "supercavity" modes in circular cylinders are a special case of the well-studied quasi-normal modes [75,126]. "Supercavity" modes are close in structure and $Q$ factor to the modes of spherical resonators of the same volume and with the same permittivity, which have the highest possible quality factors [123, 134].

*4.3.3. Perfect modes in a cylindrical nanoresonator*

In a cylindrical resonator, for any aspect ratios (and not only for exceptional ones, as in the case of "supercavity" modes), in addition to usual modes, there are also perfect non-radiating modes [40,41].

In [135], scattering spectra in a system based on a silicon dielectric cylinder were studied experimentally using highly focused axisymmetric Bessel beams with radial and azimuthal polarizations (see Fig. 22). In this work, due to the use of such polarizations, it was possible to demonstrate almost complete suppression of scattering from the Si nanodisk. Fig. 23 shows the scattered power spectrum for different polarizations.

It can be seen from Fig. 23 that for radial polarization at $\lambda$=720 nm, the scattered power practically disappears. Another interesting feature of this geometry is that the same sample, when irradiated with an azimuthally polarized beam, strongly radiates a magnetic quadrupole spherical harmonic.

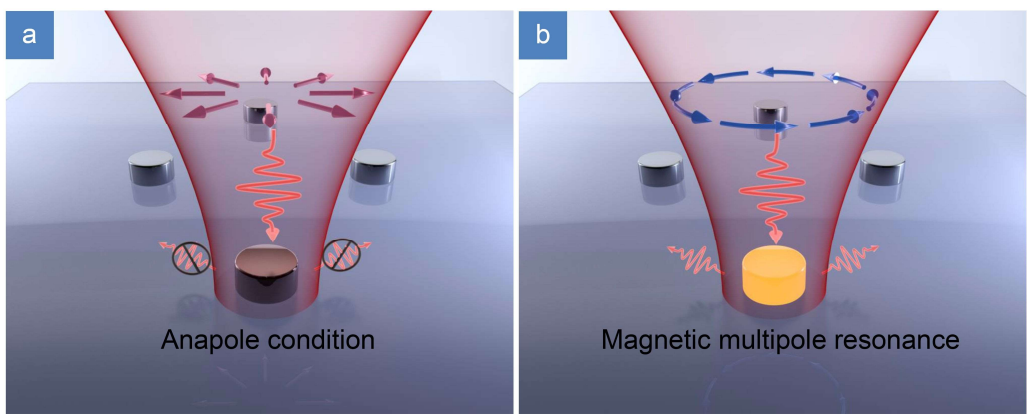

Fig.22. Excitation of a Si disk by strongly focused beams with radial (a) and azimuthal (b) polarizations [135].

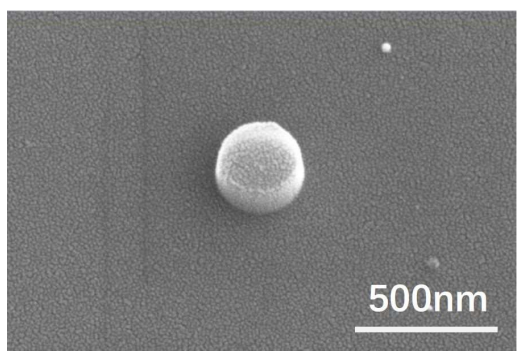
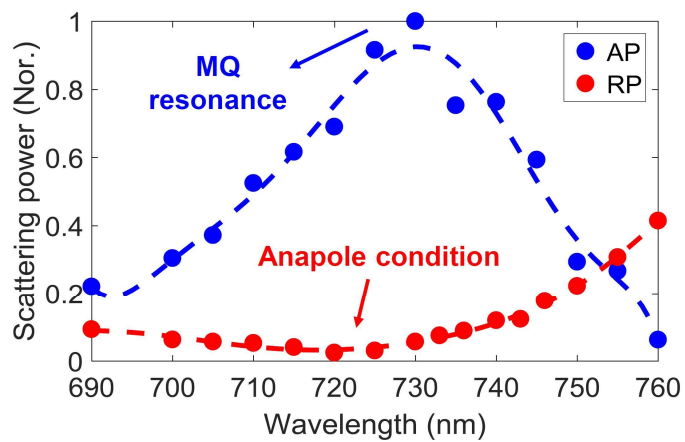

Fig.23. a) SEM image of a Si nanoresonator (*r*=150 nm, *h*=160 nm) on a SiO$_2$ substrate; b) scattered power spectrum for different polarizations of the excitation beam [135].

The authors call the disappearance of scattering at λ=720 nm the "anapole condition", but the connection of this condition with natural oscillations of the resonator is not discussed, and the "anapole condition" can correspond both to weak radiation of high-quality "supercavity" modes and to the excitation of the perfect nonradiating modes considered above [40].

The question of the existence of perfect nonradiating modes in finite-height cylinders of non-circular cross section is still open. However, perfect modes always exist for straight infinite cylinders of arbitrary cross-section [41].

## 5. Modes in nanoresonators made of metamaterials

The history of metamaterials begins with the work in UFN [136], in which V.G. Veselago showed that, unlike metals, matter (then hypothetical) with both negative permittivity and permeability allows the propagation of waves with a negative refractive index, $n = -\sqrt{\varepsilon\mu}$, that is, with oppositely directed phase and group velocities of waves.

The rapid development of this direction began with works [137,138], where metamaterials with a negative refractive index were realized in the microwave frequency range. Later, metamaterials were implemented in the IR and visible ranges (for more details see, for example, in [139,140,141]). Such materials are now called NIM (Negative Index Materials) or DNG (double negative) metamaterials. Then all substances with unusual properties began to be called metamaterials, for example, chiral metamaterials, hyperbolic metamaterials, or metamaterials with a permittivity near zero (ENZ, epsilon near zero or ZIM, zero-index metamaterials).

### 5.1. Modes in a sphere with a negative refractive index

In [142,1], the Mie theory was directly applied to describe the optical properties of nanoresonator made of a metamaterial with a negative refractive index. Fig. 24, 25 show the dependences of the amplitudes of the Mie scattering coefficients (33) and (34) as a function of permittivity and permeability.

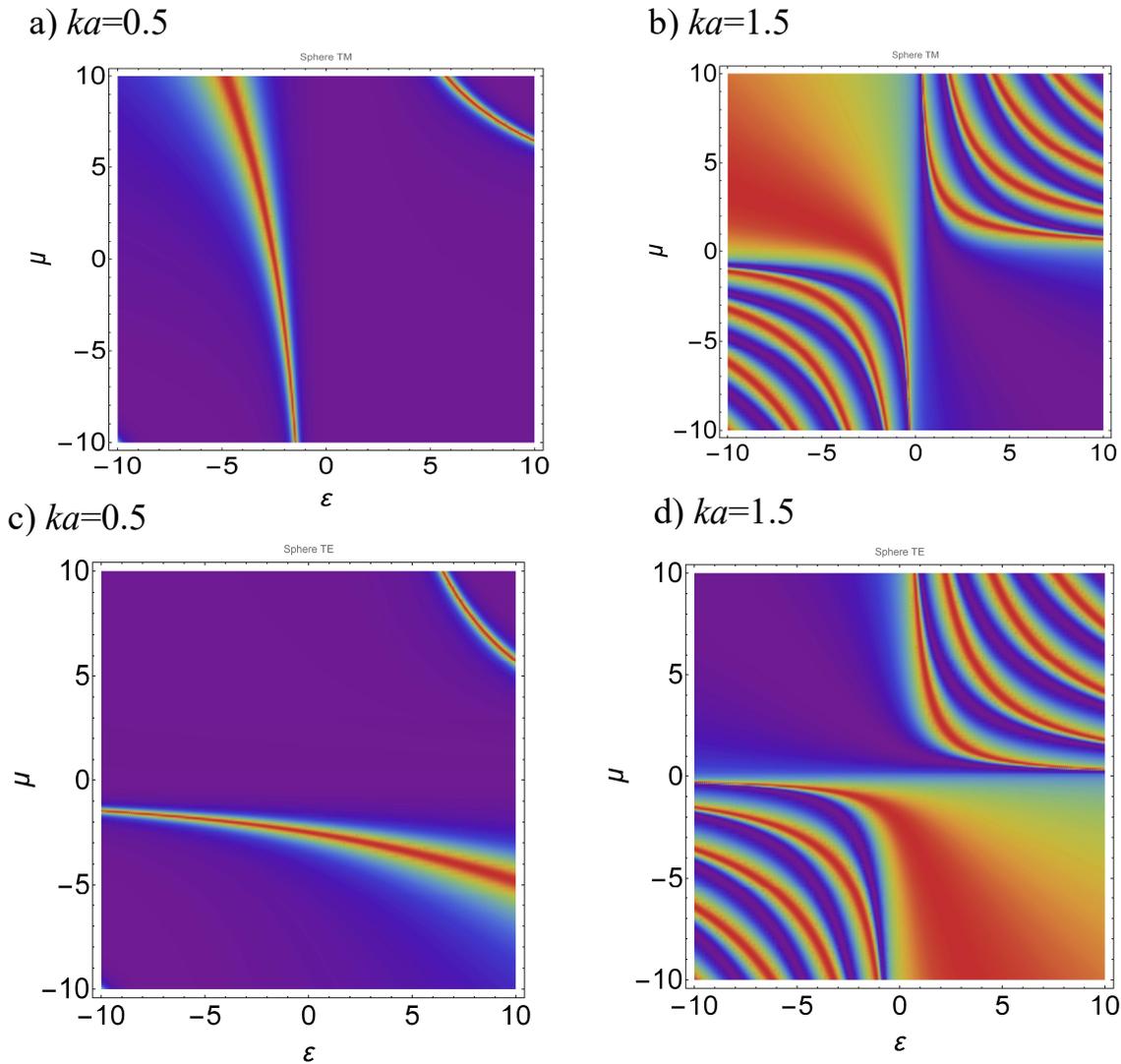

Fig. 24. The dependences of the amplitudes of the Mie scattering coefficients depending on permittivity and permeability. a),b) (TM modes) $|q_1|$ for size parameter $ka = 0.5$, $ka = 1.5$, c) d) $|p_1|$ (TE modes) for size parameter $ka = 0.5$, $ka = 1.5$. In these figures, the upper right quadrant corresponds to usual dielectrics, while the lower left quadrant corresponds to a sphere with a negative refractive index [1,142].

From Fig.24 it can be seen that the resonant properties of a sphere with a negative refractive index are only partially similar to those of a usual dielectric sphere. In it, along with the usual bulk modes, there are also surface LH (left-handed) modes, which are localized on the surface. Such a complex structure of the Mie coefficients leads to a completely different structure of modes and their quality factors compared to usual spheres.

Fig. 25 shows the dependences of the TE Mie coefficients $p_n$ on the size parameter $ka$ of the sphere.

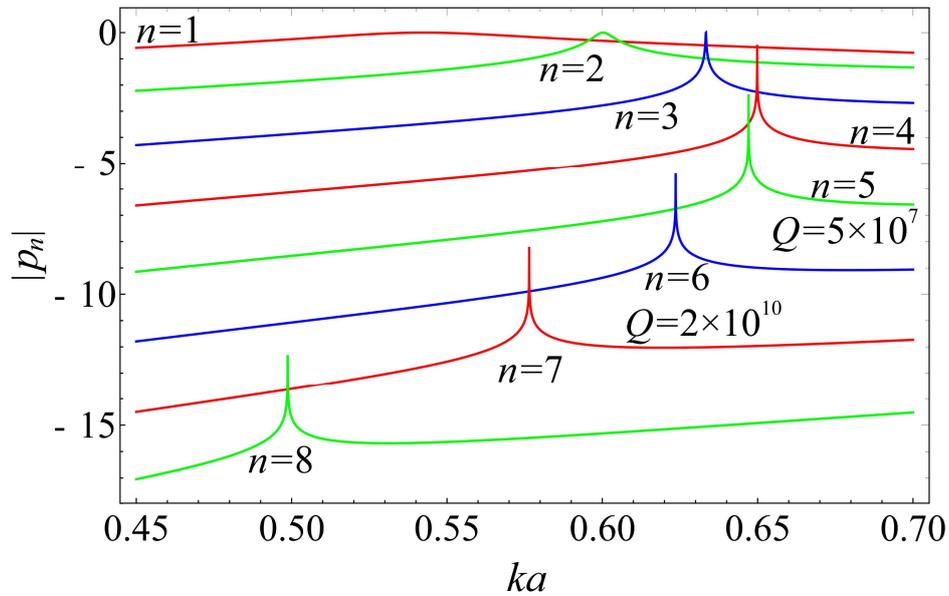

Рис.25. The dependences of the TE Mie coefficients $|p_n|$ on the size parameter $ka$ of the sphere ($\varepsilon = -15, \mu = -1.1$, $\sqrt{\varepsilon\mu} \approx -4$) [1].

Figure 25 shows clearly the highly anomalous behavior of TE resonances in a sphere made of a DNG metamaterial: with an increase in the mode multipole order, a decrease in the size parameter of the sphere $ka$ is required to achieve resonance. Moreover, the number of such resonances is finite. A decrease in the size parameter at which high multipole order modes exist leads to the fact that they become surface modes, that is, the field maximum falls on the surface of the sphere.

Fig. 26 shows the dependence of the generalized quality factor (3) of a sphere made of DNG metamaterial on the size parameter $ka$. It can be seen from this figure that, indeed, spheres with a negative refractive index have huge quality factors, which are associated with the surface nature of the LH modes. Their other most important feature is that such huge $Q$-factors arise for spheres of subwavelength sizes even at relatively small absolute values of the refractive index. Both these features distinguish modes in DNG resonators from modes in usual dielectric resonators radically.

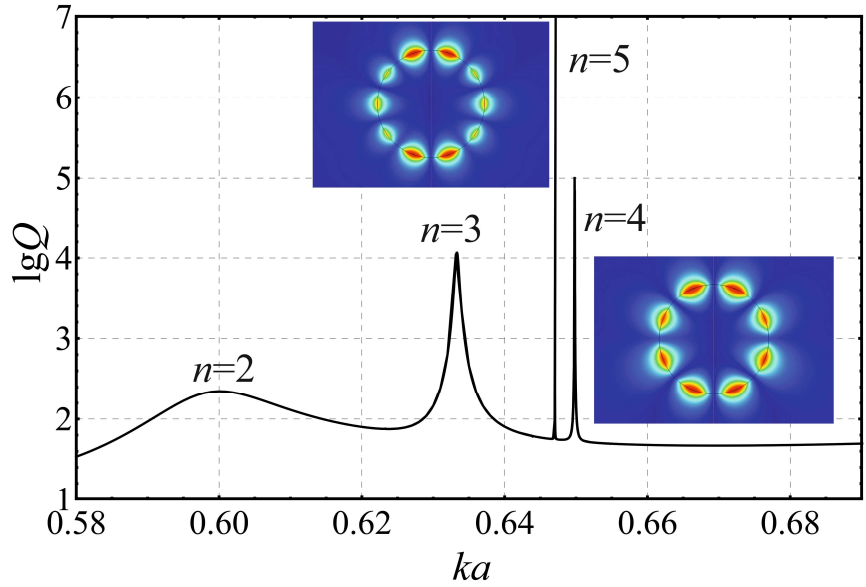

Fig.26. Dependence of the generalized quality factor (3) of TE modes in a sphere with $\varepsilon=-15$, $\mu=-1.1$ on the size parameter $ka$. The quality factors of the TE modes are $Q_2=107$, $Q_3=4125$, $Q_4=327769$, $Q_5=5.32372\times10^7$. Modes with $n=6,7,8$, clearly visible in Fig.25, are not visible here due to insufficient graphics resolution. The insets show the distribution of the electric field of the TE modes in the cross-section of the sphere [142,143,144].

5.2 Modes in a Chiral Sphere

Chiral metamaterials, where the polarization depends on both electric and magnetic fields, are related to DNG metamaterials directly. The material equations of chiral media can be expressed in the form [145,146]:

$$\mathbf{D} = \varepsilon\left(\mathbf{E} + \chi\nabla\times\mathbf{E}/k_0\right), \quad \mathbf{B} = \mu\left(\mathbf{H} + \chi\nabla\times\mathbf{E}/k_0\right), \quad (59)$$

where $\chi$ is the dimensionless chirality parameter. In bi-isotropic chiral metamaterials, where $\varepsilon, \mu$, and $\chi$ are scalar quantities, the wave vectors ($k_L, k_R$) and the refractive indices ($n_L, n_R$) for right and left polarized waves have different values:

$$k_L = k_0\frac{\sqrt{\varepsilon\mu}}{1-\chi\sqrt{\varepsilon\mu}} = k_0 n_L; \quad k_R = k_0\frac{\sqrt{\varepsilon\mu}}{1+\chi\sqrt{\varepsilon\mu}} = k_0 n_R \quad (60)$$

It can be seen from expressions (60) that for a sufficiently large $\chi\sqrt{\varepsilon\mu}$ one of the refractive indices ($n_R$ or $n_L$) will necessarily become negative. In [147], it was proposed to use this fact to implement media with a negative refractive index. Chiral media can be implemented in several ways [148-152]. Fig. 27 shows possible implementations of a chiral spherical nanoresonator.

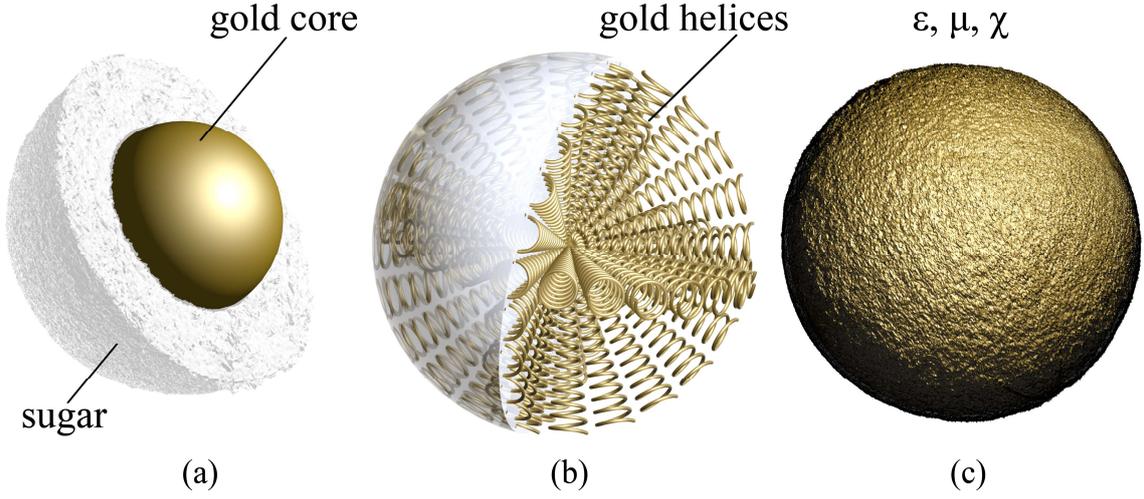

Fig. 27. Possible implementation of a chiral sphere: a) a gold nanosphere covered with a shell of sugar; (b) radial spirals forming a sphere; (c) effective description of a chiral nanoparticle.

It was shown in [153,154] within the framework of the quasi-static approximation that instead of a dipole plasmon resonance, $\varepsilon(\omega)+2=0$, a chiral-plasmon dipole resonance arises in chiral spheres, the condition for which is

$$(\varepsilon+2)(\mu+2)=4\varepsilon\mu\chi^2 \qquad (61)$$

A complete electrodynamic theory of resonances in chiral spherical nanoparticles of arbitrary sizes was developed in [155-157], where expressions for analogues of the Mie coefficients were found. These expressions are much more complicated, since in chiral spheres it is no longer possible to separate the modes into TE and TM ones. However, it is possible to introduce more complex modes, called in [156] modes of $A$ and $B$ types. Figure 28 shows the dependence of the scattering coefficient $T_4^A$ (type $A$ mode), which reduces to the Mie coefficient for TM mode scattering, $q_4$, at $\chi=0$, on the permittivity $\varepsilon$ and permeability $\mu$ for various values of the size parameter $k_0 a$ and chirality $\chi$.

It can be seen from this figure that the structure of the modes of the chiral sphere is much more complicated than in the case of the DNG sphere, since here the chiral-plasmon resonance (61) arises additionally, manifesting itself in a discontinuity in the dispersion curve of the surface DNG modes.

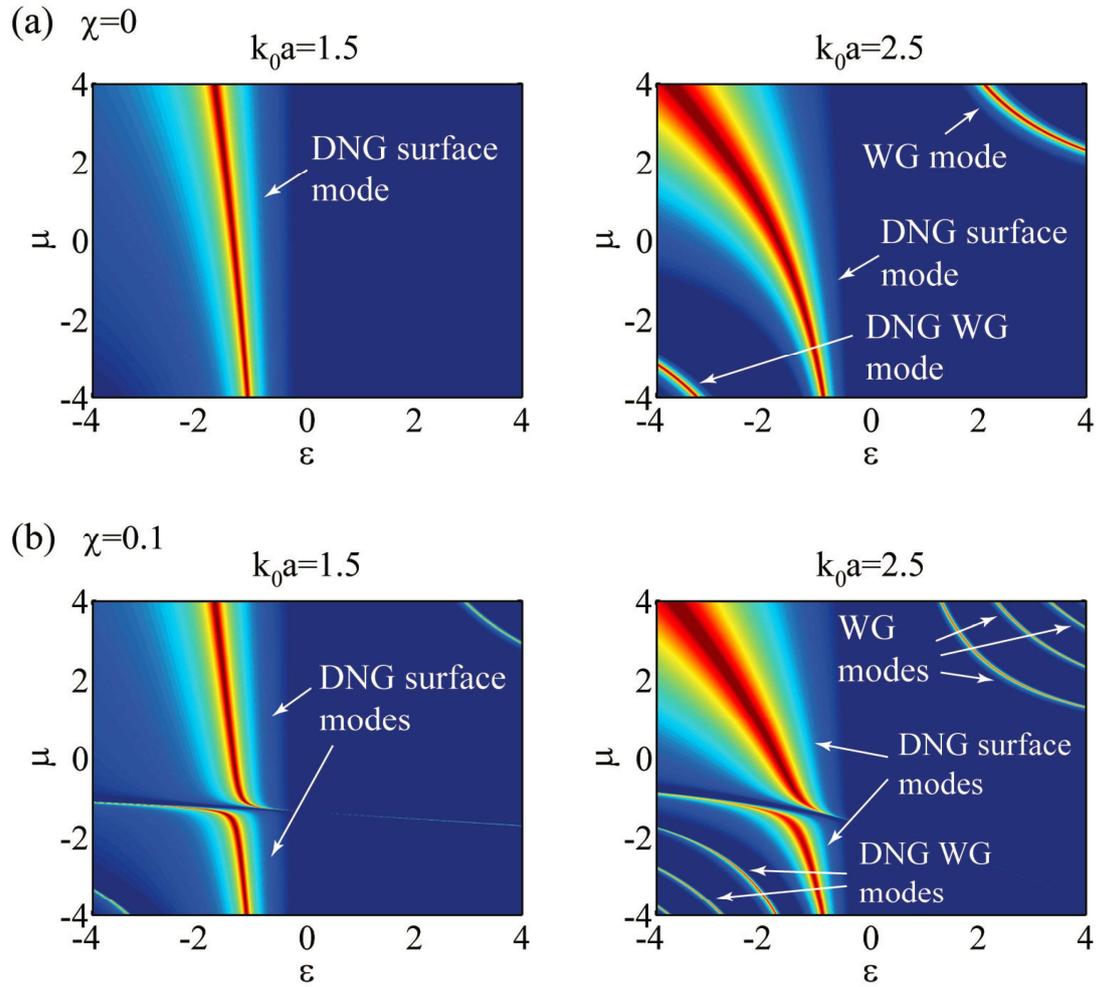

Fig. 28. The dependence of the absolute values of the scattering coefficient of *A* modes, $T_4^A$, of a chiral spherical particle on the permittivity $\varepsilon$ and the permeability $\mu$ for various values of the size parameter $k_0 a$. (a) $\chi = 0$; (b) $\chi = 0.1$. The nanosphere is located in vacuum [156].

The dispersion equation that determines the eigenfrequencies of the chiral sphere has the form [155,156]

$$\Delta_n = W_n(L)V_n(R) + W_n(R)V_n(L) = 0,$$
$$W_n(J) = (\varepsilon/\mu)^{1/2} \psi_n(k_J a) \zeta_n^{(1)\prime}(k_0 a) - \psi_n'(k_J a) \zeta_n^{(1)}(k_0 a), \qquad (62)$$
$$V_n(J) = \psi_n(k_J a) \zeta_n^{(1)\prime}(k_0 a) - (\varepsilon/\mu)^{1/2} \psi_n'(k_J a) \zeta_n^{(1)}(k_0 a),$$

where $\psi_n(z) = z j_n(z)$ and $\zeta_n^{(1)}(z) = z h_n^{(1)}(z)$ are the Ricatti-Bessel functions, the prime denotes the derivative, the index *J* takes the values *L, R* (see (60)). It is important that, as in a usual dielectric sphere, the expression for the determinant $\Delta_n$ does not depend on the azimuthal quantum number *m*. Figure 29 shows the projection of the dependence of the eigenfrequencies of the chiral sphere with ε = 2 + 0.04i on χ, Re$k_0a$, Im$k_0a$ onto the Re$k_0a$, Im$k_0a$ plane for an orbital quantum number $n = 1$.

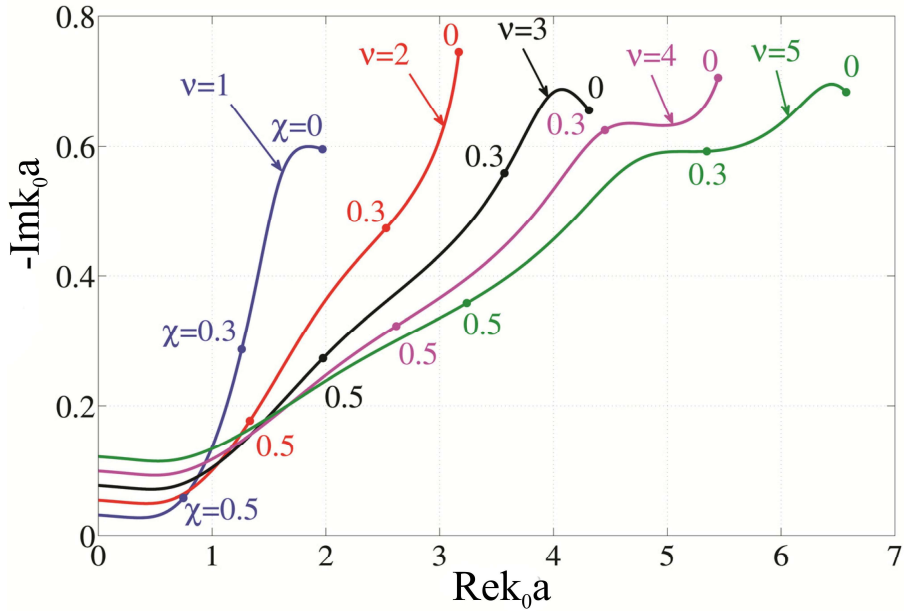

Fig. 29. Projection of the dependence of the eigenfrequencies of the chiral sphere with ε = 2 + 0.04i on χ, Re$k_0a$, Im$k_0a$ onto the Re$k_0a$, Im$k_0a$ plane for an orbital quantum number $n = 1$. Different colors of the curves correspond to different radial quantum numbers ν = 1...5. Colored labels near points indicate the chirality parameter χ at that point (3rd coordinate) [157].

It can be seen from this figure that an increase in chirality (moving downwards along the curves) leads to a significant increase in the quality factor of natural oscillations. This is mainly due to the fact that the wave number of one of the polarizations in a chiral medium can increase indefinitely (the pole in (60)), and therefore the wavelength of such waves in a sphere is much smaller than in vacuum, leading to an effective increase in the size of the sphere and the quality factor of the modes existing in it.

Figure 30 shows the distribution of the z-component of the electric field in the plane z = 0 (the shaded plane in the geometry inset) for different values of the dimensionless chirality parameter corresponding to different values of the radial quantum number ν = 1, 2, 3, 4 when the sphere is excited by a plane wave.

Other regimes of eigenoscillations for a chiral sphere were considered in [156-158]. Natural oscillations in a cluster of chiral spheres [159] and in a chiral sphere with an asymmetric shell [160] were also studied in detail.

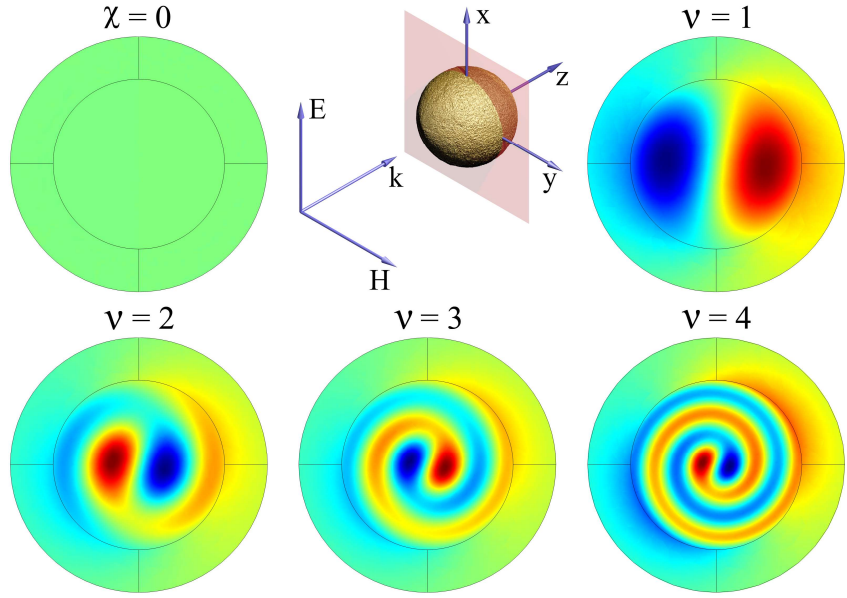

Fig. 30. Geometry of the problem of the scattering of a plane wave with a $\lambda=570$ nm chiral sphere $\varepsilon = 2 + 0.04i$ and $a = 70$ nm and spatial distribution of the z-component of the electric field $E_z$ in the $z = 0$ plane (the shaded plane on geometry inset) for different values of dimensionless chirality parameter corresponding to different values of the radial quantum number $\nu = 1$ ($\chi \approx 0.5$), $\nu = 2$ ($\chi \approx 0.6$), $\nu = 3$ ($\chi \approx 0.65$), $\nu = 4$ ($\chi \approx 0.67$), eigenmodes with $n = 1$, $m = 1$. The case $\chi = 0$ is also presented for comparison.

### 5.3. Modes in a Sphere Made of Hyperbolic Metamaterial

In [161], modes were considered in a spherical nanoresonator made of a hyperbolic metamaterial (HMM), that is, an anisotropic material where the diagonal values of the permittivity tensor have different signs [162] (Fig. 31 a).

As is known [162], the key feature of the HMM is the possibility of the existence of propagating waves with an unlimited radial wave number (Fig.31 (b) ), enabling the existence of resonant modes in spheres of arbitrarily small sizes (in the approximation of a homogeneous medium, of course) ( see Fig. 31 c, d). With a decrease in the outer radius of the resonator $R$, the Q-factors of the modes increase unlimitedly, according to the law $Q_{rad} \propto R^{-(2n+1)}$, where $n$ is the order of the multipole moment of the mode. Although the authors of Ref. [161] say that the high-$Q$ modes found by them can be attributed to whispering gallery modes, this is apparently not the case, and it can be seen from Fig. 37 c, d that these modes are concentrated at the center of the resonator.

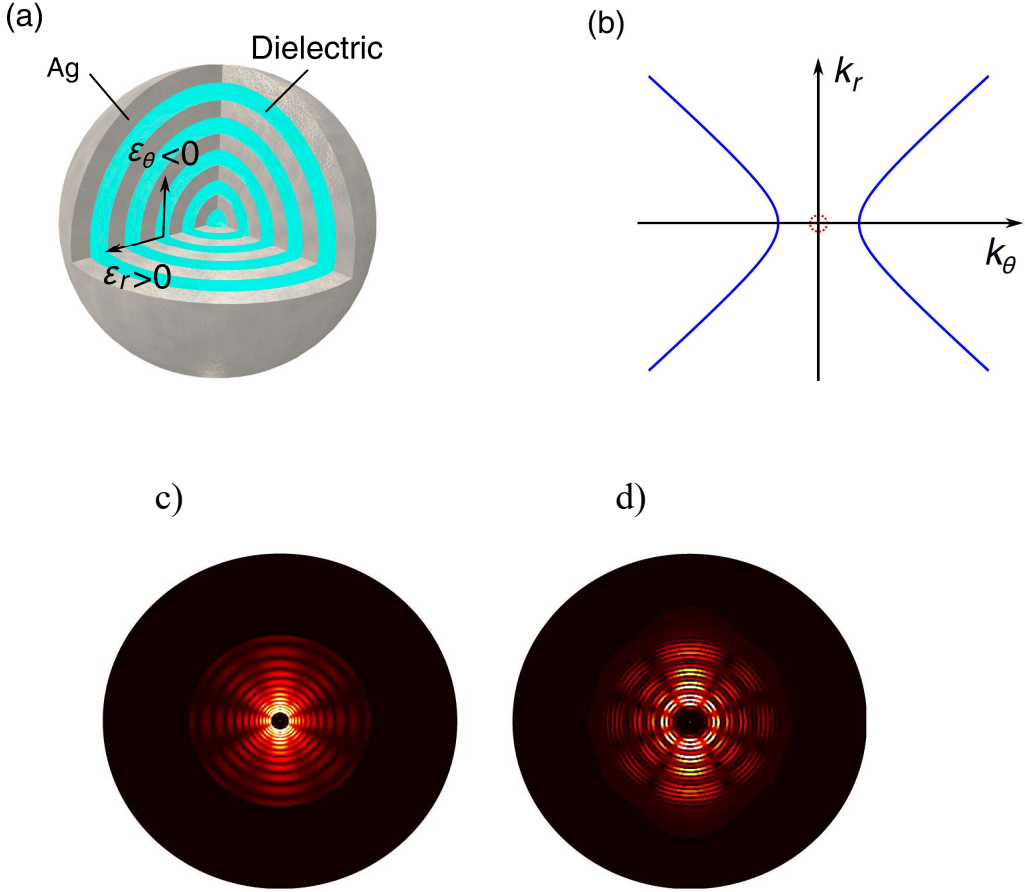

Fig.31. (a) Geometry of a hyperbolic resonator consisting of alternating metal and dielectric layers. (b) HMM isofrequency contour in momentum space calculated in the effective medium approximation with $\lambda = 1000$ nm. The small dotted circle in the center represents the isofrequency contour in vacuum at the same wavelength. (c) The field distribution in the TM resonant mode with an orbital number $n=4$ in a homogeneous hyperbolic resonator and in a multilayer structure (d). In the case (c), the HMM resonator has the inner radius of 20 nm and the outer radius of 200 nm and the resonant wavelength $\lambda=992$ nm. The HMM resonator in case (d) consists of 20 functional layers 10 nm thick each, the resonant wavelength in (d) is $\lambda = 977$ nm [161].

From a mathematical point of view, this is because the radial distributions of TM fields inside a spherical HMM resonator are described not by usual Bessel functions of a half-integer order, but by Bessel functions with an imaginary order [163]:

$$J_{l+1/2}\left(\sqrt{\varepsilon}k_0 r\right) \Rightarrow J_{g(l)+1/2}\left(\sqrt{\varepsilon_\theta}k_0 r\right), \qquad (63)$$

where the imaginary order is given by

$$g(l)+\frac{1}{2}=\frac{1}{2}\sqrt{1+4\frac{l(l+1)\varepsilon_\theta}{\varepsilon_r}}, \varepsilon_\theta<0, \varepsilon_r>0 \qquad (64)$$

The Bessel functions with imaginary order do not tend to zero at the origin leading to the concentration of fields at the center of the sphere.

The properties of HMM resonators are generally similar to those of resonators made of DNG or chiral metamaterials: in all cases, modes can exist for arbitrarily small external dimensions of the resonators. From a physical point of view, these resonators are fundamentally different: in nanoresonators made of hyperbolic and chiral metamaterials, the modes are bulk, while in DNG nanoresonators, superhigh-$Q$ modes are of a surface nature.

## 6. Examples of Possible Applications of Optical Nanoresonators

Although optical nanoresonators have very interesting fundamental properties, their wide practical application is yet to come. Nevertheless, the principles of operation of several optical nanodevices based on the resonant properties of nanoparticles have already been experimentally demonstrated.

1) First of all, the attempts to create sources of coherent radiation - nanolasers and spasers, both based on plasmonic nanoresonators [2,4], and on the basis of dielectric nanoresonators [22, 25,26, 164] with the use of optical nanoresonators should be noted. There are also proposals to use HMM nanoresonators for nanolasers [165].

2) Optical nanoresonators can also be used effectively to control the spontaneous emission of elementary quantum systems [143,144,166-172] and to construct bright artificial fluorophores [91] and quantum photonic circuits [173] on this basis (where a 1D photonic crystal nanoresonator was used).

3) An important area of application of optical nanoresonators is to increase the efficiency of solar cells, and for this purpose, not only plasmonic, but also dielectric nanoresonators are used [174-178].

4) The high sensitivity of natural oscillations in optical nanoresonators to the refractive index of the environment makes it possible to create highly sensitive biosensors on this basis [110,179-185].

5) The high field concentration in high-$Q$ modes of optical nanoresonators and the correct choice of pump polarization increase the efficiency of generation of the second [21, 30,188,189] and the third [29,190,191,192] harmonics and other nonlinear effects [17, 21, 29] significantly (by 2 orders of magnitude compared to the best microresonators [186,187]).

6) Electromagnetic resonances in dielectric nanoresonators can be used to implement optical magnetism ($\mu \neq 1$) [120,193,194].

## 7. Conclusion

The review presents the current state of the rapidly developing field of optics - the optics of 3D nanoresonators, that is, resonators with subwavelength dimensions in all directions. In this case, the main attention is paid to the fundamental aspects of describing the optical properties of such nanoresonators. The rapid development of this direction by the efforts of various scientific groups led to discovery of several new effects and to the emergence of bright new terms related to them, which are not precisely defined often. This complicates the further development of this direction, and we try to connect the new terms with each other and with generally accepted definitions.

Mainly, the review considers highly symmetric nanoresonators without internal losses, which have ultrahigh $Q$-factors at resonant frequencies. However, a whole class of nanoresonators - asymmetric nanoresonators - remained outside the scope of the review. It is difficult to expect superhigh quality factors in the usual sense of the word from such resonators, but interesting non-stationary field distributions can be realized in them, which can serve as the basis for information processing on a single chip and other applications. Super scattering regimes can also exist in asymmetric nanocavities.

The materials of nanoresonators have a fundamentally important effect on their properties, and therefore both plasmonic and dielectric nanoresonators are considered in the review. The review also considers nanoresonators made from hypothetical metamaterials. Although these metamaterials have not yet been realized at the nanoscale (see, however, [195]), the extreme properties of resonators made from them make the search for such metamaterials extremely topical.

In the review, modes of natural oscillations are considered more or less detailed, but the methods for their effective excitation are only outlined. At present, optical nanoresonators are mainly excited by an external (far) field, and therefore, studies aimed at creating far-field configurations that effectively interact with eigenfields are extremely important. Especially interesting for excitation of symmetrical nanoresonators are Bessel beams with a certain (radial or azimuthal) polarization. For example, only the use of an azimuthally polarized beam instead of a linearly polarized one makes it possible to increase the efficiency of the second harmonic generation by 2 orders of magnitude [30]. Even more important from a practical point of view is the excitation of natural oscillations in nanoresonators using near fields for the construction of photonic nanochips. It is natural to consider elementary quantum systems (dye molecules or quantum dots) as the sources of near fields. Such studies are just beginning, and mostly in theory since the experiments are very complex.

We believe that fruitful achievements and great breakthroughs will be made in this area in the future, and we hope that this review will give an additional impetus to research in this direction.

**Acknowledgments:**

The reported study was funded by RFBR, project number 20-12-50136.